\documentclass[aip, pof,graphicx]{revtex4-1}

\usepackage{graphicx}
\usepackage{dcolumn}
\usepackage{bm}
\usepackage[utf8]{inputenc}
\usepackage[T1]{fontenc}
\usepackage{mathptmx}
\usepackage{color}
\usepackage{longtable}
\draft 

\begin{document}


\title[Crystallization and jamming in narrow fluidized beds]{Crystallization and jamming in narrow fluidized beds\\
This article may be downloaded for personal use only. Any other use requires prior permission of the author and AIP Publishing. This article appeared in Phys. Fluids 32, 083303 (2020) and may be found at https://doi.org/10.1063/5.0015410.} 



\author{Fernando David C\'u\~nez}
\author{Erick M. Franklin}%
 \email{franklin@fem.unicamp.br}
 \thanks{Corresponding author}
\affiliation{ 
School of Mechanical Engineering, UNICAMP - University of Campinas,\\
Rua Mendeleyev, 200, Campinas, SP, Brazil
}%


\date{\today}

\begin{abstract}
A fluidized bed is basically a suspension of granular material by an ascending fluid in a tube, and it has a rich dynamics that includes clustering and pattern formation. When the ratio between the tube and grain diameters is small, different behaviors can be induced by high confinement effects. Some unexpected and curious behaviors, that we investigate in this paper, are the crystallization and jamming of grains in liquids with velocities higher than that for incipient fluidization, supposed to maintain the grains fluidized. In our experiments, performed in a vertical tube of transparent material, different grains, water velocities, resting times, and velocity decelerations were used. An analysis of the bed evolution based on image processing shows that, after a decreasing flow that reaches a velocity still higher than that for incipient fluidization, grains become organized in lattice structures of high compactness, where they are trapped though with small fluctuations. These structures are initially localized and grow along time, in a similar manner as happens in phase transitions and glass formation. After a certain time, if the liquid velocity is slightly increased, jamming occurs, with grains being completely blocked and their fluctuation disappearing. We show that different lattice structures appear depending on the grain type. Our results provide new insights into fluidization conditions, glass-like formation and jamming.
\end{abstract}


\maketitle 

\section{INTRODUCTION}
\label{sec:intro}

A fluidized bed is basically a suspension of granular material by an ascending fluid, and it is frequently employed in industry for its high rates of mass and heat transfers between the fluid and solids. From the mechanical point of view, fluidized beds have a rich dynamics, with shocks between grains, slipping, contacts, and shedding of small vortices occurring at the grain scale, and the formation of contact networks, clusters, plugs and other structures at larger scales. The coexistence of a large number of important mechanisms at different scales makes the problem complex, and, depending if the employed fluid is a gas or a liquid, some mechanisms have more or less relative importance. For instance, in the case of solid-liquid fluidized beds (SLFBs), fluid drainage dissipates a considerable amount of energy, shocks transmit less energy (although they are of importance), and virtual mass force may be important \cite{Cunez2}. 

Narrow SLFBs, typically with thickness between 10 to 100 grain diameters, have been studied for their distinct behavior when compared to usual beds \cite{Anderson, ElKaissy, Didwania, Zenit, Zenit2, Duru2, Duru, Aguilar, Ghatage}. Different instabilities and patterns have been identified for that case, such as the propagation of transverse waves \cite{Duru2} and bubbles \cite{Duru}, for instance. In the very narrow case, for which the ratio of the tube diameter $D$ to that of grains $d$ is less than 10, different behaviors are induced by high confinement effects \cite{Cunez, Cunez2}. C\'u\~nez and Franklin \cite{Cunez} investigated both experimentally and numerically the structure of a very narrow bed ($D/d$ = 4.2), and found the formation of alternating regions of high and low particle fractions, called plugs and bubbles, respectively. They showed, by analyzing the network of contact forces, that those structures are associated with the presence of close walls. The same kind of structure is observed in bidesperse beds \cite{Cunez2}, their presence being more important as $D/d$ decreases. In the case of water, C\'u\~nez and Franklin \cite{Cunez2} showed that the added mass force has an effect of decreasing by roughly 10\% the characteristic time of the narrow-bed dynamics.

Goldman and Swinney \cite{Goldman} investigated experimentally the crystallization and jamming in a SLFB with $D/d$ of the order of 100. Crystallization consists in the organization of the bed in a static lattice of high compactness. In this lattice, grains have no macroscopic motion, but maintain small fluctuations, usually referred to as microscopic motion. The reasons for the appearance of crystallization are still not understood, but it occurs more frequently during partial defluidization, i.e., by decreasing the fluid velocity of an already fluidized bed until a velocity slightly above that of minimum fluidization, $U_{mf}$, is reached. Jamming usually appears when a crystallized bed is forced upwards by submitting it to a fluid velocity slightly higher than that of crystallization. Goldman and Swinney \cite{Goldman} showed that crystallization occurs during partial defluidization, and that the final state depends on the decreasing rate, mainly because grains sediment for deceleration rates above a threshold value. They showed that bed crystallization has similarities with equilibrium glass transition, one of them being the rate dependence and another one the initially localized solid-like regions that grow along time. They showed also that once crystallization is complete, grains lose their macroscopic motion while preserving the microscopic one, and that if the liquid velocity is slightly increased the system is completely jammed.

Tariot et al. \cite{Tariot} and Gauthier et al. \cite{Gauthier} investigated experimentally the compaction of granular material in a SLFB of rectangular cross section with thickness of 10$d$. In the experiments, the authors fluidized the bed and then let the grains settle by turning off the water flow. Afterward, they imposed a water flow and measured the evolution of bed compactness for different water velocities, using two types of flows: a continuous water velocity below $U_{mf}$, and pulses (bursts) of water below \cite{Gauthier} and above \cite{Tariot, Gauthier} $U_{mf}$. They found a slow increase in bed compactness for both cases, with the compaction of the pulsed cases reaching higher values when compared to that of continuous flow. Gauthier et al. \cite{Gauthier} showed that under continuous flow compaction is more effective for velocities approaching $U_{mf}$, and for pulsed flows compaction is much more effective for velocities above $U_{mf}$. Finally, Gauthier et al. \cite{Gauthier} proposed $d/U_{mf}$ as a timescale and that the time evolution of particle fraction follows a logarithmic law.

Most of previous studies dealing with hindrance of fluidization on narrow or small beds concerned either $U_{mf}$ or compaction/expansion of beds from a macroscopic point of view \cite{Jin, Rao, Nascimento, Li2}. For example, Jin et al. \cite{Jin} investigated the expansion and collapse of beds in a pulsed SLFB by analyzing particle concentrations, while Rao et al. \cite{Rao} investigated the effects of bed height and diameter on $U_{mf}$ by measuring pressure gradients and using a continuous model (which is suitable for large beds). Li et al. \cite{Li2} and do Nascimento et al. \cite{Nascimento} investigated micro SLFBs, with $D$ of the order of 1 mm and $d$ of the orfder of 10-100 $\mu$m, and found that, given the small scales of grains, hindrances are caused by superficial forces acting on grains, including relatively strong adhesion forces. None of them investigated how grains are organized during compaction, the appearance of crystallization within the bed, its susceptibility to jamming, or the behavior of very narrow beds ($D/d$ $<$ 5).

Very few studies were devoted to crystallization and jamming in SLFBs and, although some physical aspects of these processes were investigated previously, many questions remain to be clarified. For instance, the involved scales and the effect of higher confinement remain to be understood. In the present paper, we investigate experimentally the crystallization and jamming of grains in very narrow SLFBs, where we used $D/d$ = 4.2 and $D/d$ = 3.2 and different grain densities, water velocities, resting times and velocity decelerations. We show that different confinement and grain types generate distinct lattice structures that are initially localized and grow along time. The present results provide new insights into fluidization conditions, glass-like formation and jamming.

\section{EXPERIMENTAL SETUP}
\label{sec:setup}

\begin{figure}[ht]
	\centering
	\includegraphics[width=0.8\columnwidth]{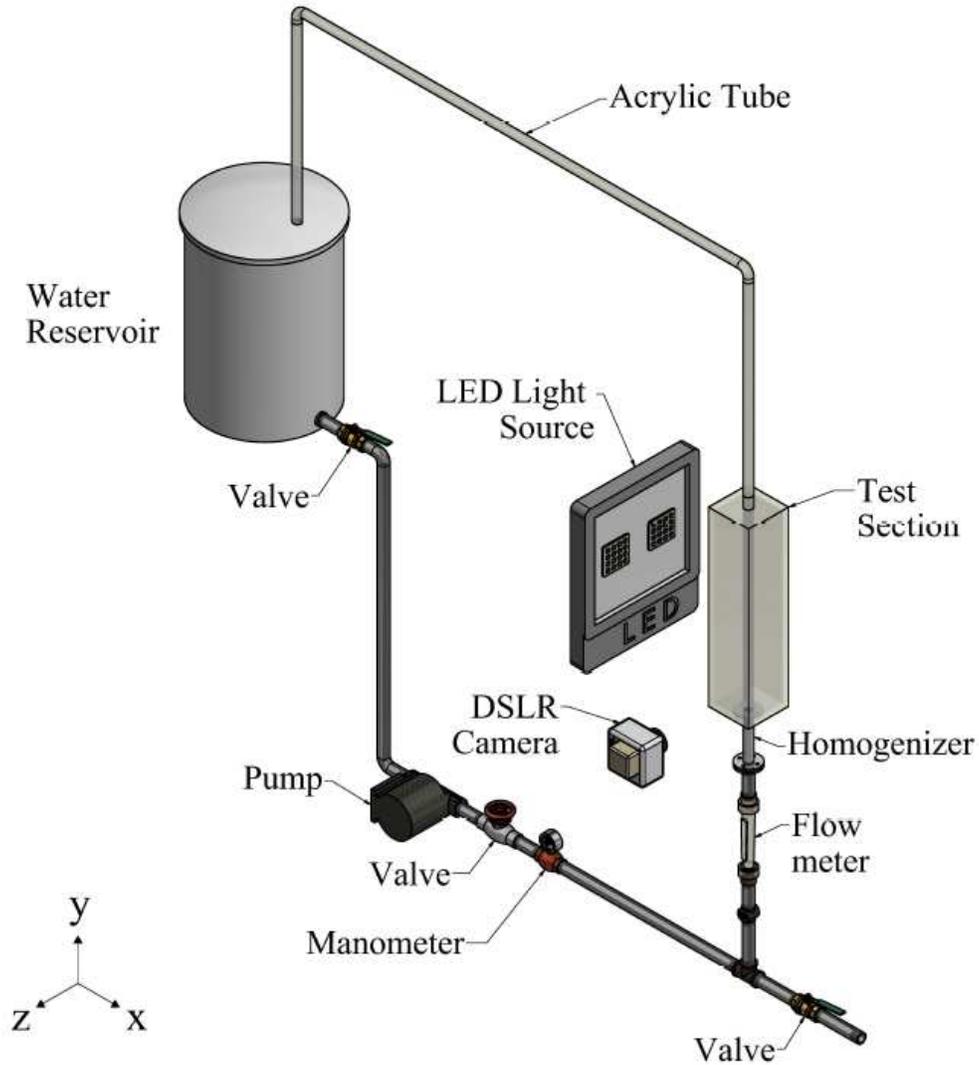}
	\caption{Layout of the experimental setup.}
	\label{fig:1}
\end{figure}

\begin{figure}[ht]
	\centering
	\includegraphics[width=0.35\columnwidth]{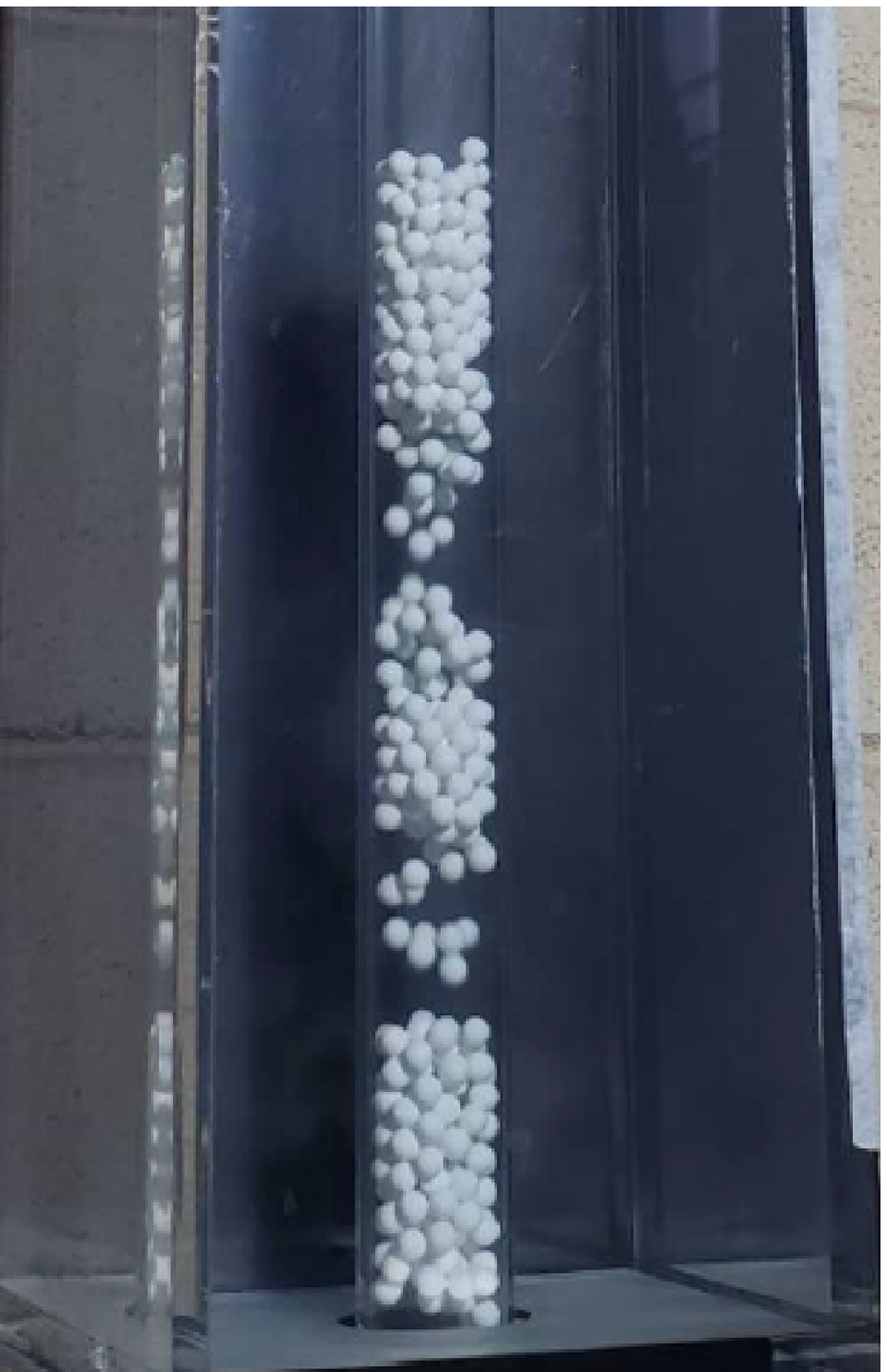}
	\caption{Fluidized bed occupying part of the test section.}
	\label{fig_test_section}
\end{figure}

The experimental setup consisted of a water tank with a heat exchanger, a centrifugal pump, a flow meter, a flow homogenizer, a 1.2-m-long tube of transparent material (polymethyl methacrylate - PMMA) oriented vertically, and a return line, where water flowed in closed loop following the order just described. The vertical tube was aligned vertically within $\pm 3^{\circ}$ and had an internal diameter of 25.4 mm. The test section corresponded to 0.65 m in length from the entrance of the tube, and a visual box filled with water was placed around it to minimize parallax distortions. Figure\ref{fig:1} shows the layout of the experimental setup and Fig. \ref{fig_test_section} shows a photograph of part of the test section.

Controlled grains were settled in the test section, forming a granular bed, and upward water flows were imposed by controlling the rotation of the centrifugal pump, which had a maximum flow capacity of 4100 l/h. The flow homogenizer was a 150-mm-long tube containing packed beads with $d$ = 6 mm between fine wire screens and it was placed just upstream the test section. In our tests, water temperatures were within 25$^{\circ}$C $\pm$ 3$^{\circ}$C.

A camera of complementary metal-oxide-semiconductor (CMOS) type was placed perpendicularly to the test section in order to acquire images of the beds. The camera resolution was of 1920 px $\times$ 1080 px at 60 Hz and it was branched to a computer system that controlled both the camera and the pump rotation. With that, the velocity variations were automated and synchronized with the camera. The camera frequency was set to 30 Hz and the region of interest (ROI) to 1920 px $\times$ 211 px in all our tests. We used a lens of $60$ mm focal distance and F2.8 maximum aperture mounted on the camera, and we ended with images where 1 px corresponded to approximately 0.1 mm. Lamps of light emission diode (LED) were branched to a continuous-current source to provide a stable light.

\begin{figure}[h!]
	\centering
	\includegraphics[width=0.5\columnwidth]{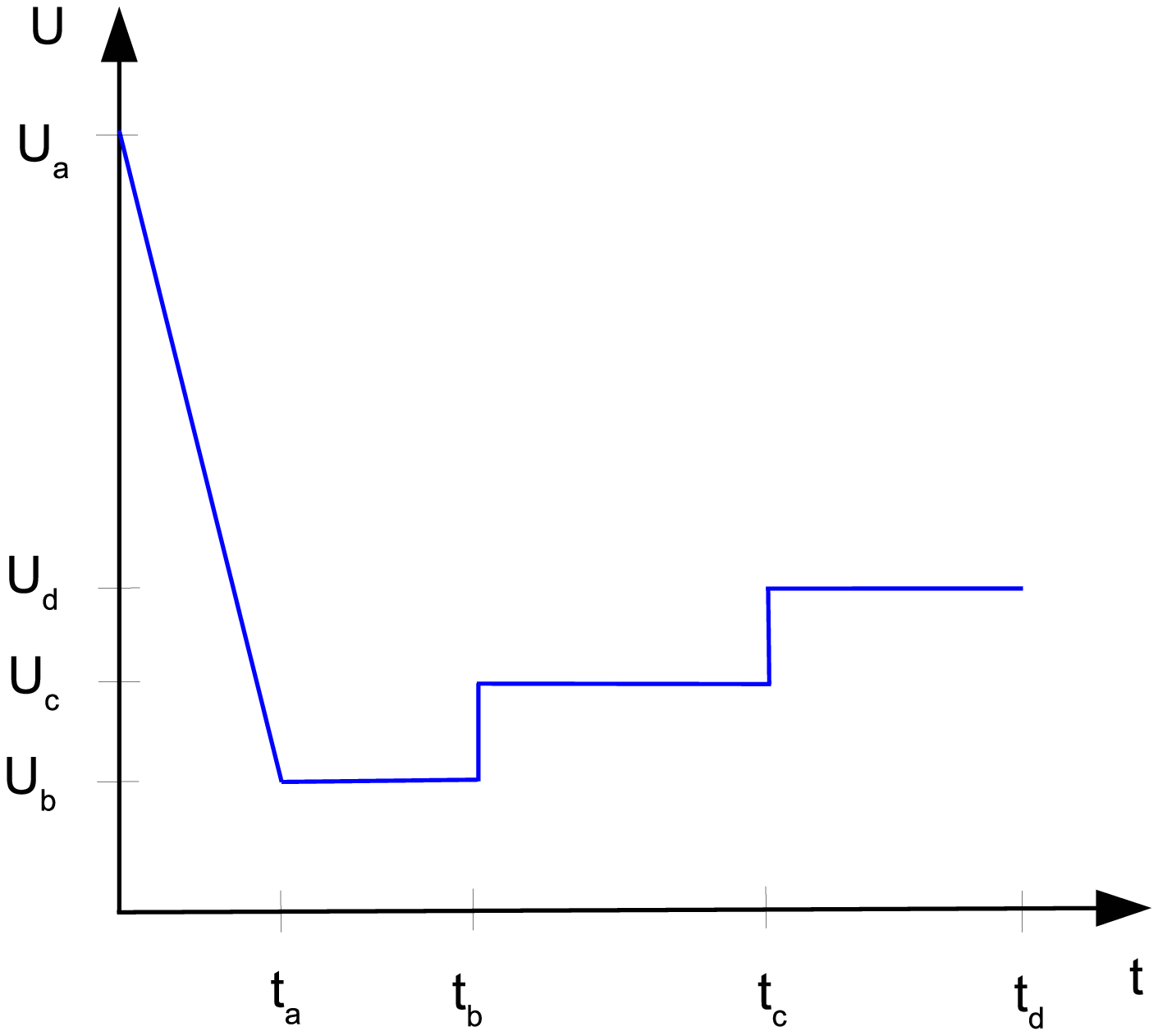}
	\caption{Scheme of the imposed water velocities along tests. Values are listed in Tabs. \ref{table:table1} and \ref{table:table2}.}
	\label{fig_water velocities}
\end{figure}

\begin{table}[h!]
\caption{Grain diameter $d$, particle type, density ratio $S$, terminal Reynolds number $Re_t$, terminal Stokes number $St_t$, bed height at the inception of fluidization $h_{if}$, particle fraction at the inception of fluidization $\phi_0$, settling velocity $v_s$, incipient fluidization velocity $U_{if}$, and velocities $U_{a}$, $U_{b}$, $U_{c}$ and $U_{d}$ normalized by $U_{if}$.}
\label{table:table1}
\centering
\begin{tabular}{c c c c c c c c c c c c c}  
\hline\hline
$d$ & Species & $S$ & $Re_t$  & $St_t$ & $h_{if}$ & $\phi_0$ & $v_s$ & $U_{if}$ & $U_a/U_{if}$ & $U_b/U_{if}$ & $U_c/U_{if}$ & $U_d/U_{if}$\\
mm & $\cdots$ & $\cdots$ & $\cdots$ & $\cdots$ & m & $\cdots$ & m/s & m/s & $\cdots$ &$\cdots$ & $\cdots$ & $\cdots$\\ [0.5ex] 
\hline 
6.0 & 1 & 3.69 & 4026 & 1650 & 0.174 & 0.51 & 0.118 & 0.118 & 1.86 & 1.02 & 1.07 & 1.16\\
6.0 & 2 & 2.50 & 2967 & 824 & 0.165 & 0.54 & 0.076 & 0.076 & 2.89 & 1.01 & 1.08 & 1.37\\
8.0 & 3 & 2.50 & 4654 & 1293 & 0.220 & 0.48 & 0.120 & 0.120 & 1.83 & 1.01 & 1.05 & 1.14\\
\hline
\hline 
\end{tabular}
\end{table}

Three different beds were investigated, the first consisting of 400 alumina beads with $d_1$ = 6 mm and $S_1$ = $\rho_p / \rho_f$ = 3.69, the second of 400 glass beads with $d_2$ = 6 mm and $S_2$ = $\rho_p / \rho_f$ = 2.5, and the third of 200 glass beads with $d_3$ = 8 mm and $S_3$ = $\rho_p / \rho_f$ = 2.5, where $\rho_p$ is the density of the bead material and $\rho_f$ the density of the fluid. Microscopy images of the used grains are available in the Supplementary Material. For these beds, the ratios between the tube and grain diameters were of $D/d_1$ = 4.2, $D/d_2$ = 4.2 and $D/d_3$ = 3.2, and the numbers of Stokes $St_t \,=\, v_t d \rho_p / (9\mu_f)$ and Reynolds $Re_t \,=\, \rho_f  v_t d / \mu_f$ based on terminal velocities, where $v_t$ is the terminal velocity of one single particle and $\mu_f$ is the dynamic viscosity of the fluid, are shown in Tab. \ref{table:table1}. Values of $St_t$ and $Re_t$ in our tests indicate that grains have considerable inertia with respect to the employed fluid. Table \ref{table:table1} presents also, for the inception of fluidization, the bed heights $h_{if}$, the particle fraction $\phi_0$, and the settling velocity $v_s$, the latter computed based on the Richardson--Zaki correlation, $v_s = v_t\phi_0^{2.4}$. Because we had optical access to the bed, we determined the inception of fluidization based on acquired images by detecting the motion of grains and associating the respective water velocity with an incipient fluidization velocity $U_{if}$. This procedure is different from the usual method for determining $U_{mf}$ based on pressure differences. While, on the one hand, beds consisting of a small number of grains are susceptible to relatively high pressure variations at the inception of fluidization, on the other hand it is relatively easy to detect the motion of their grains and verify that they are suspended. The values of $U_{if}$, obtained from image processing and not necessarily equal to $U_{mf}$, are also shown in Tab. \ref{table:table1}.

In our tests, we initially fluidized the bed by imposing a cross-sectional mean velocity $U_a$ = 0.219 m/s, corresponding to $U_a/U_{if1}$ = 1.86, $U_a/U_{if2}$ = 2.89 and $U_a/U_{if3}$ = 1.83, during 90 s, then we reduced the velocity at a given deceleration rate until reaching a value $U_{b}$ slightly above that of incipient fluidization ($\approx$ 101 to 102\% $U_{if}$), which was maintained for a certain time. Afterward, we increased the velocity slightly, to $U_{c}$ around 105 to 108\% $U_{if}$, and maintained it for a given time, and finally we increased it slightly again to $U_{d}$ around 114 to 137\% $U_{if}$ and maintained it for a given time interval. For each grain type, different velocity decelerations and time intervals were used. Figure \ref{fig_water velocities} shows the imposed water velocities, and Tab. \ref{table:table1} presents the values of velocities and Tab. \ref{table:table2} the time instants of each test.

\begin{longtable}[c]{c c c c c c c c c c c c}
\caption{Test number, particle type, programed times $t_a$, $t_b$, $t_c$ and $t_d$, time of the beginning of crystallization $t_{cry}$, number of times that crystallization occurred $N_{cry}$, intensity of crystallization $I_{cry}$ = $\Delta t_{cry}/(t_b - t_a)$, number of times that jamming occurred $N_{jam}$, intensity of jamming $I_{jam}$ = $\Delta t_{jam}/\Delta t_2$, and intensity of the volcano-like structure $I_{vol}$ = $\Delta t_{vol}/\Delta t_3$.}\label{table:table2}\\

 \hline
 \hline
Test & Species & $t_a$ & $t_b$  & $t_c$ & $t_d$ & $t_{cry}$ & $N_{cry}$  & $I_{cry}$ & $N_{jam}$ & $I_{jam}$ & $I_{vol}$\\
$\cdots$ & $\cdots$ & s & s & s & s & s & $\cdots$ & $\cdots$ & $\cdots$ & $\cdots$ & $\cdots$\\ [0.5ex] 
 \hline
 \endfirsthead

 \hline
 \multicolumn{12}{c}{Continuation of Table \ref{table:table2}}\\
 \hline
Test & Species & $t_a$ & $t_b$  & $t_c$ & $t_d$ & $t_{cry}$ & $N_{cry}$  & $I_{cry}$ & $N_{jam}$ & $I_{jam}$ & $I_{vol}$\\
$\cdots$ & $\cdots$ & s & s & s & s & s & $\cdots$ & $\cdots$ & $\cdots$ & $\cdots$ & $\cdots$\\ [0.5ex] 
 \hline
 \endhead

 \hline
 \endfoot

 \hline
 \hline\hline
 \endlastfoot

1 & 1 & 10 & 600 & 1200 & 1500 & 485 & 1 & 0.19 & 1 & 1.00 & $\cdots$\\
2 & 1 & 10 & 900 & 1800 & 2100 & 830 & 1 & 0.08 & 1& 0.61 & $\cdots$\\
3 & 1 & 10 & 1200 & 2400 & 2700 & $\cdots$ & 0 & 0.00 & 0 & 0.00 & $\cdots$\\
4 & 1 & 10 & 2400 & 4800 & 5100 & 1950 & 1 & 0.19 & 1 & 0.00 & $\cdots$\\
5 & 1 & 10 & 4800 & 9600 & 9900 & 2400 & 2 & 0.40 & 1 & 0.03 & $\cdots$\\
6 & 1 & 40 & 600 & 1200 & 1500 & 300 & 1 & 0.54 & 1 & 0.04 & $\cdots$\\
7 & 1 & 40 & 900 & 1800 & 2100 & 310 & 2 & 0.45 & 1 & 1.00 & $\cdots$\\
8 & 1 & 40 & 1200 & 2400 & 2700 & 300 & 1 & 0.28 & 0 & 0.00 & $\cdots$\\
9 & 1 & 40 & 2400 & 4800 & 5100 & 330 & 1 & 0.55 & 2 & 0.28 & $\cdots$\\
10 & 1 & 40 & 4800 & 9600 & 9900 & 320 & 3 & 0.55 & 2 & 0.55 & $\cdots$\\
11 & 1 & 80 & 600 & 1200 & 1500 & $\cdots$ & 0 & 0.00 & 1 & 0.20 & $\cdots$\\
12 & 1 & 80 & 900 & 1800 & 2100 & 620 & 1 & 0.34 & 1 & 1.00 & $\cdots$\\
13 & 1 & 80 & 1200 & 2400 & 2700 & 800 & 1 & 0.36 & 2 & 0.11 & $\cdots$\\
14 & 1 & 80 & 2400 & 4800 & 5100 & $\cdots$ & 0 & 0.00 & 1 & 0.00 & $\cdots$\\
15 & 1 & 160 & 4800 & 9600 & 9900 & 1140 & 2 & 0.57 & 0 & 0.00 & $\cdots$\\
16 & 1 & 160 & 600 & 1200 & 1500 & $\cdots$ & 0 & 0.00 & 0 & 0.00 & $\cdots$\\
17 & 1 & 160 & 900 & 1800 & 2100 & 780 & 1 & 0.16 & 1 & 0.24 & $\cdots$\\
18 & 1 & 160 & 1200 & 2400 & 2700 & 1140 & 1 & 0.06 & 1 & 0.01 & $\cdots$\\
19 & 1 & 160 & 2400 & 4800 & 5100 & 2250 & 1 & 0.07 & 2 & 0.42 & $\cdots$\\
20 & 1 & 160 & 4800 & 9600 & 9900 & 2080 & 1 & 0.59 & 1 & 0.01 & $\cdots$\\
21 & 1 & 200 & 600 & 1200 & 1500 & $\cdots$ & 0 & 0.00 & 0 & 0.00 & $\cdots$\\
22 & 1 & 200 & 900 & 1800 & 2100 & $\cdots$ & 0 & 0.00 & 0 & 0.00 & $\cdots$\\
23 & 1 & 200 & 1200 & 2400 & 2700 & 1140 & 1 & 0.06 & 1 & 0.08 & $\cdots$\\
24 & 1 & 200 & 2400 & 4800 & 5100 & 700 & 1 & 0.77 & 1 & 0.00 & $\cdots$\\
25 & 1 & 200 & 4800 & 9600 & 9900 & 2310 & 1 & 0.54 & 1 & 0.02 & $\cdots$\\
26 & 2 & 12 & 600 & 1200 & 1500 & 30 & 1 & 0.97 & 1 & 1.00 & 0.00\\
27 & 2 & 12 & 1200 & 2400 & 2700 & 35 & 1 & 0.98 & 0 & 0.00 & 0.77\\
28 & 2 & 12 & 4800 & 9600 & 9900 & 20 & 1 & 1.00 & 2 & 0.76 & 0.27\\
29 & 2 & 60 & 600 & 1200 & 1500 & 70 & 1 & 0.98 & 1 & 0.48 & 0.00\\
30 & 2 & 60 & 1200 & 2400 & 2700 & 80 & 1 & 0.98 & 1 & 0.55 & 0.37\\
31 & 2 & 60 & 4800 & 9600 & 9900 & 80 & 1 & 1.00 & 1 & 0.75 & 0.60\\
32 & 2 & 240 & 600 & 1200 & 1500 & 280 & 1 & 0.89 & 1 & 1.00 & 0.17\\
33 & 2 & 240 & 1200 & 2400 & 2700 & 275 & 1 & 0.96 & 2 & 0.75 & 0.72\\
34 & 2 & 240 & 400 & 9600 & 9900 & 270 & 1 & 0.99 & 1 & 1.00 & 0.07\\
35 & 2 & 300 & 600 & 1200 & 1500 & 350 & 1 & 0.83 & 2 & 0.85 & 0.23\\
36 & 2 & 300 & 1200 & 2400 & 2700 & 340 & 1 & 0.96 & 2 & 0.60 & 0.27\\
37 & 2 & 300 & 4800 & 9600 & 9900 & 332 & 1 & 0.99 & 1 & 1.00 & 0.12\\
38 & 3 & 50 & 600 & 1500 & 1800 & 90 & 1 & 0.93 & 1 & 1.00 & $\cdots$\\
39 & 3 & 50 & 1200 & 2700 & 3000 & 96 & 1 & 0.96 & 1 & 1.00 & $\cdots$\\
40 & 3 & 50 & 4800 & 9900 & 10200 & 120 & 1 & 0.99 & 1 & 1.00 & $\cdots$\\

\end{longtable}

\section{RESULTS AND DISCUSSION}
\label{sec:results}

A general picture of our experimental observations can be summarized as follows. After decreasing the water flow from the velocity $U_{a}$ to $U_{b}$, for different deceleration times $t_a$, the velocity $U_{b}$, still higher than $U_{if}$, was sustained for an interval $t_b - t_a$. In these conditions, for most of cases, grains became organized in a lattice structure filling the tube cross section, where grains were trapped though with small fluctuations (microscopic motions \cite{Goldman}). These structures were initially localized and grew along time, in a similar manner as happens in phase transitions and glass formation, and in some cases they appeared and disappeared more than once. The time for the beginning of crystallization, $t_{cry}$, the number of times crystallization appeared, and their intensity, defined as $I_{cry}$ = $\Delta t_{cry}/(t_b - t_a)$, where $\Delta t_{cry}$ is the duration of crystallization, are shown in Tab. \ref{table:table2}. After $\Delta t_1$ = $t_b $ $-$ 0, the liquid velocity was increased to $U_{c}$ and was sustained for an interval $\Delta t_2$ = $t_c - t_b$. Under these conditions, for most of cases, jamming occurred, where grains became completely blocked and their microscopic motion disappeared. As with crystallization, jamming appeared and disappeared more than once in some instances. The number of times that jamming appeared, $N_{jam}$, and their intensity, defined as $I_{jam}$ = $\Delta t_{jam}/\Delta t_2$, where $\Delta t_{jam}$ is the duration of jamming, are presented in Tab. \ref{table:table2}. Finally, at $t_c$ the water velocity was increased to $U_{d}$ and was maintained for $\Delta t_3$ = $t_d - t_c$. For species 1, the bed fluidized again, but for species 2 a new crystallization occurred, with grains being organized in a volcano-like lattice, where grains were organized in a static structure in contact with the tube wall while absent in the center of the tube cross section, forming an annular structure. The intensity of the volcano-like structure, defined as $I_{vol}$ = $\Delta t_{vol}/\Delta t_3$, where $\Delta t_{vol}$ is the duration of the volcano, is also shown in Tab. \ref{table:table2}. For species 3, the bed remained jammed and it was necessary a higher velocity to unjam it (not shown here). The bed structure and the behavior of individual grains are analyzed and discussed next. Movies of one of our experiments are available as Supplementary Material, and movies and images of several experiments, as well as the numerical scripts used for image processing, are available in Mendeley Data \cite{Supplemental3}.

\subsection{Macroscopic observations}

\begin{figure}[h!]
	\centering
	\includegraphics[width=0.99\columnwidth]{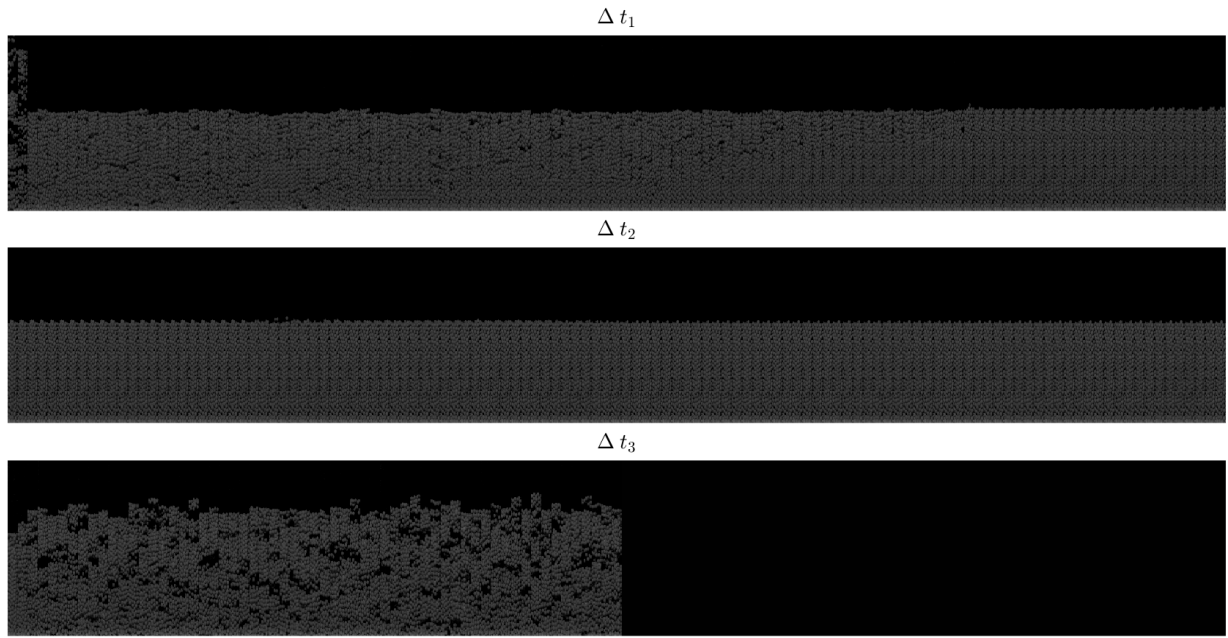}
	\caption{Snapshots placed side by side for Test 1. The first row corresponds to $\Delta t_1$, the second to $\Delta t_2$ and the third to $\Delta t_3$. Time between frames is of 5 s.}
	\label{fig_test1}
\end{figure}

\begin{figure}[h!]
	\centering
	\includegraphics[width=0.99\columnwidth]{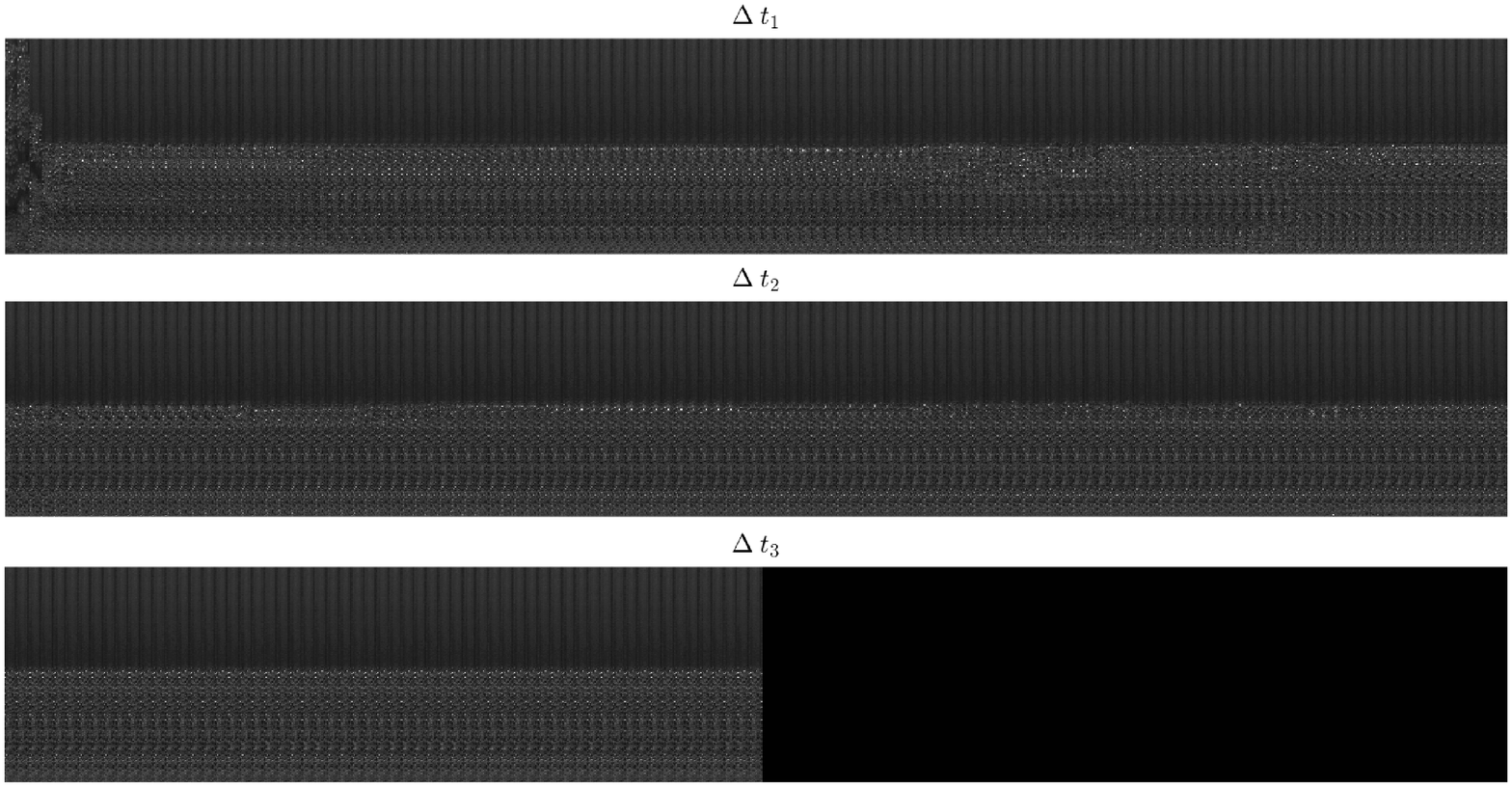}
	\caption{Snapshots placed side by side for Test 26. The first row corresponds to $\Delta t_1$, the second to $\Delta t_2$ and the third to $\Delta t_3$. Time between frames is of 5 s.}
	\label{fig_test26}
\end{figure}

\begin{figure}[h!]
	\centering
	\includegraphics[width=0.99\columnwidth]{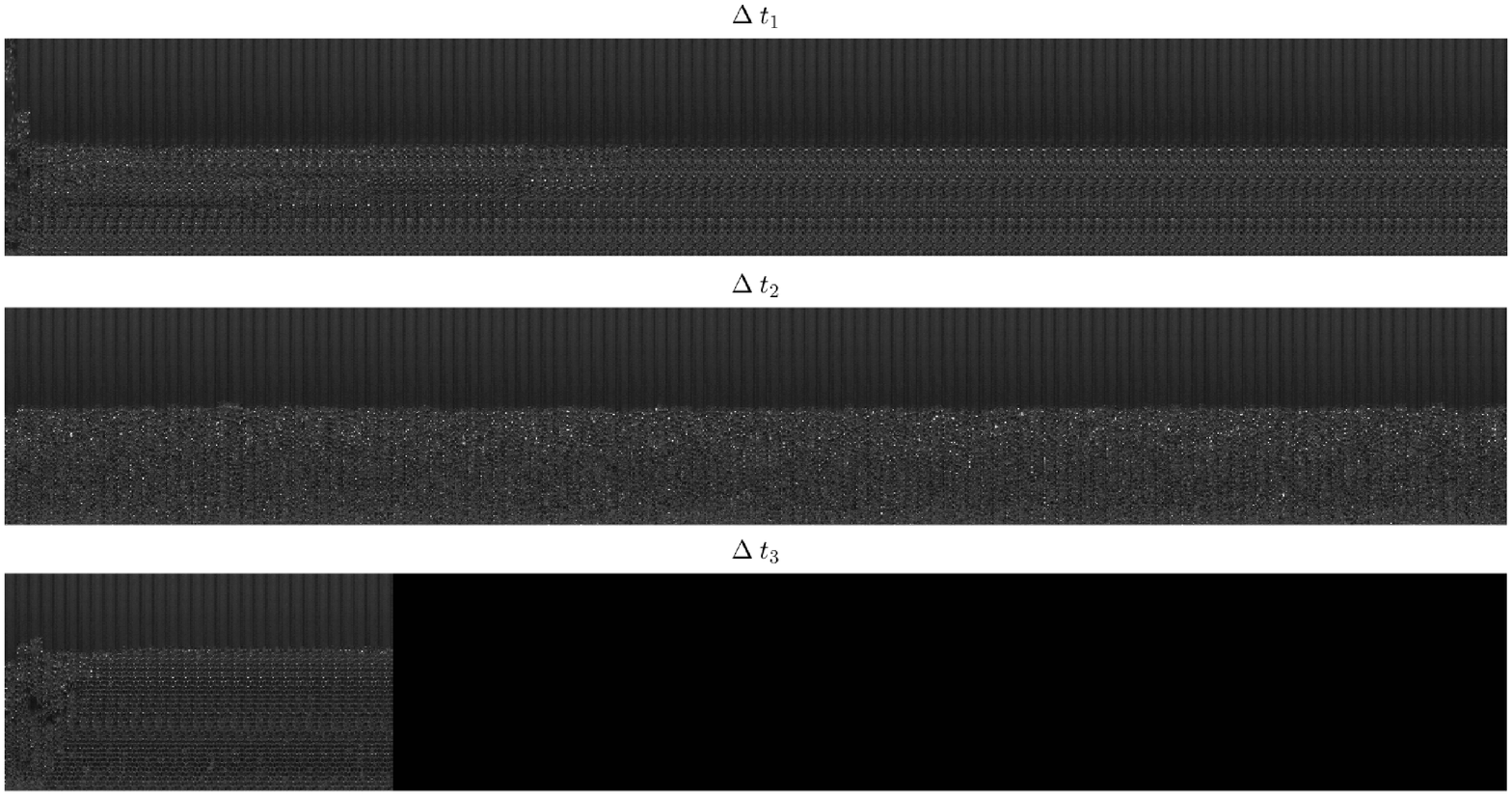}
	\caption{Snapshots placed side by side for Test 27. The first row corresponds to $\Delta t_1$, the second to $\Delta t_2$ and the third to $\Delta t_3$. Time between frames is of 10 s.}
	\label{fig_test27}
\end{figure}

We investigated the evolution of the bed structure by comparing consecutive movie frames along time, for three stages of bed evolution, namely $\Delta t_1$, $\Delta t_2$ and $\Delta t_3$. In order to visualize macroscopically the different bed structures, we placed sequences of snapshots side by side, as can be seen in Figs. \ref{fig_test1}, \ref{fig_test26} and \ref{fig_test27} for tests 1, 26 and 27, respectively (Tab. \ref{table:table2}). These figures consist basically in spatiotemporal plots without filtering, since they are constructed directly with unprocessed images. In Figs. \ref{fig_test1} to \ref{fig_test27}, the time between frames was increased in order to enable them to fit the page while showing macroscopically the bed structures. 

\begin{figure}[h!]
	\centering
	\includegraphics[width=0.99\columnwidth]{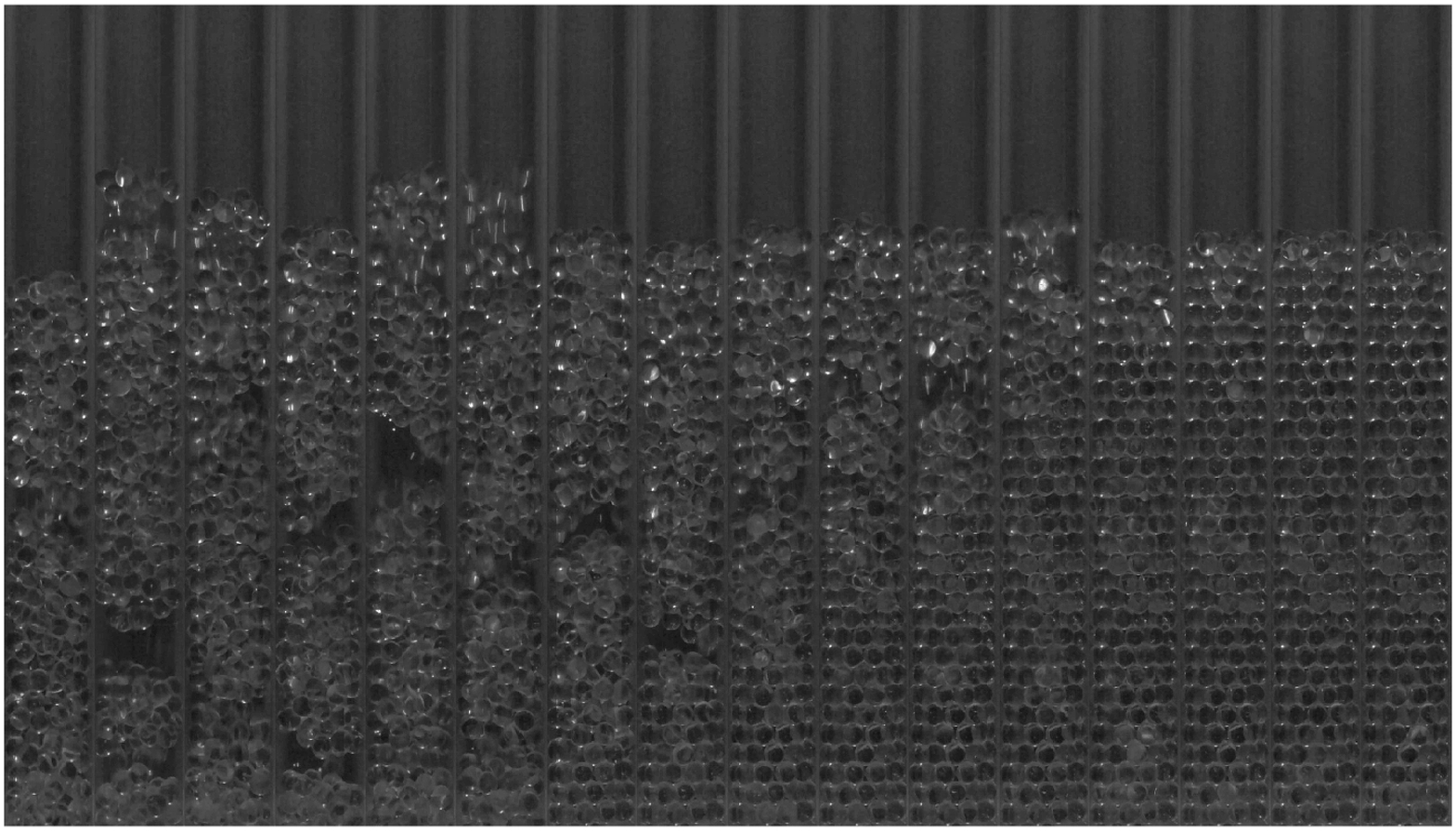}
	\caption{Snapshots placed side by side for the intervals of Test 27 in which the volcano-like structure is formed. Time between frames is of 5 s.}
	\label{fig_test27_zoom}
\end{figure}

In Figs. \ref{fig_test1} to \ref{fig_test27} we observe initially a decrease in the bed height due to the decelerating flow, with the occurrence of crystallization after the velocity $U_b$ was attained. During $\Delta t_2$, with $U_{c}$ $>$ $U_{b}$, we observe jamming in Figs. \ref{fig_test1} and \ref{fig_test26}, with grains maintaining their respective positions along frames, but not in Fig. \ref{fig_test27}. Finally, during $\Delta t_3$ we observe fluidization again in Fig. \ref{fig_test1}, jamming in Fig. \ref{fig_test26}, and a volcano-like structure in Fig. \ref{fig_test27}. The formation of the volcano-like structure is shown in more detail in Fig. \ref{fig_test27_zoom}, and can be better observed in one of the movies available as Supplementary Material \cite{Supplemental3}, corresponding to test 29 (Tab. \ref{table:table2}), and in snapshots of other cases, available in Mendeley Data \cite{Supplemental3}.

\begin{figure}[h!]
   \begin{minipage}[c]{0.35\linewidth}
    \begin{center}
     \includegraphics[width=.99\linewidth]{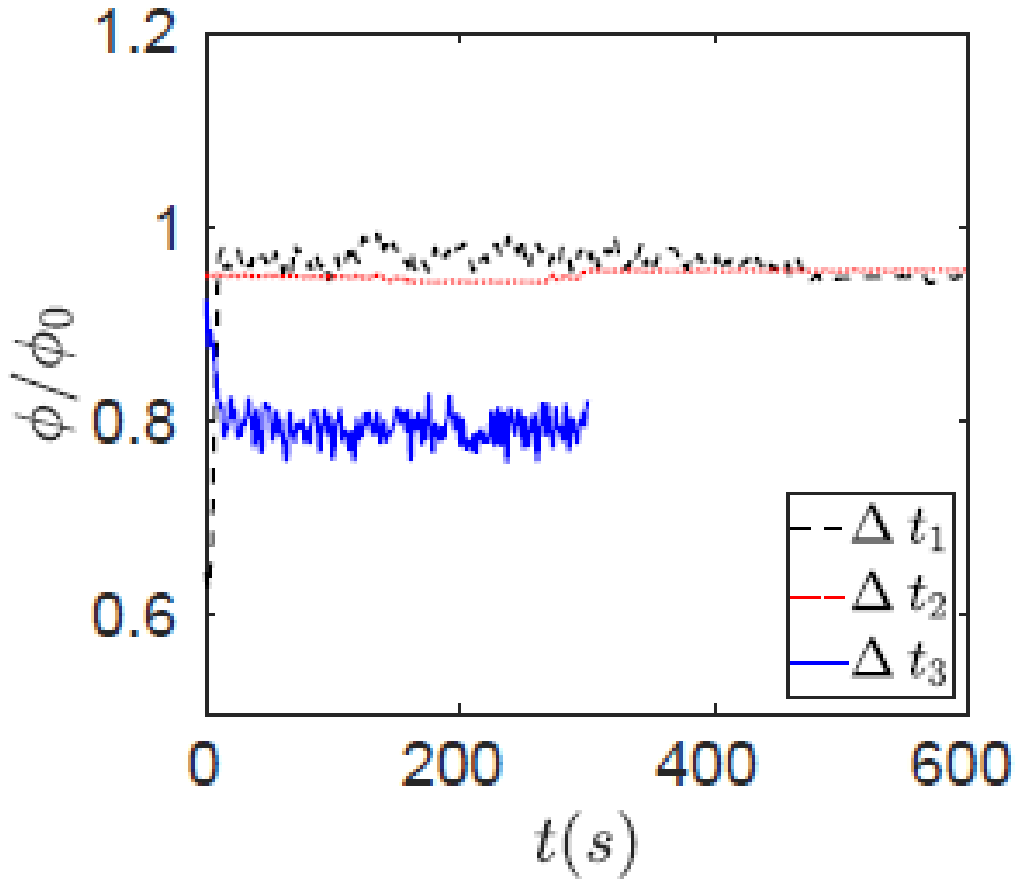}\\
		(a)
    \end{center}
   \end{minipage}
   \begin{minipage}[c]{0.35\linewidth}
    \begin{center}
      \includegraphics[width=.99\linewidth]{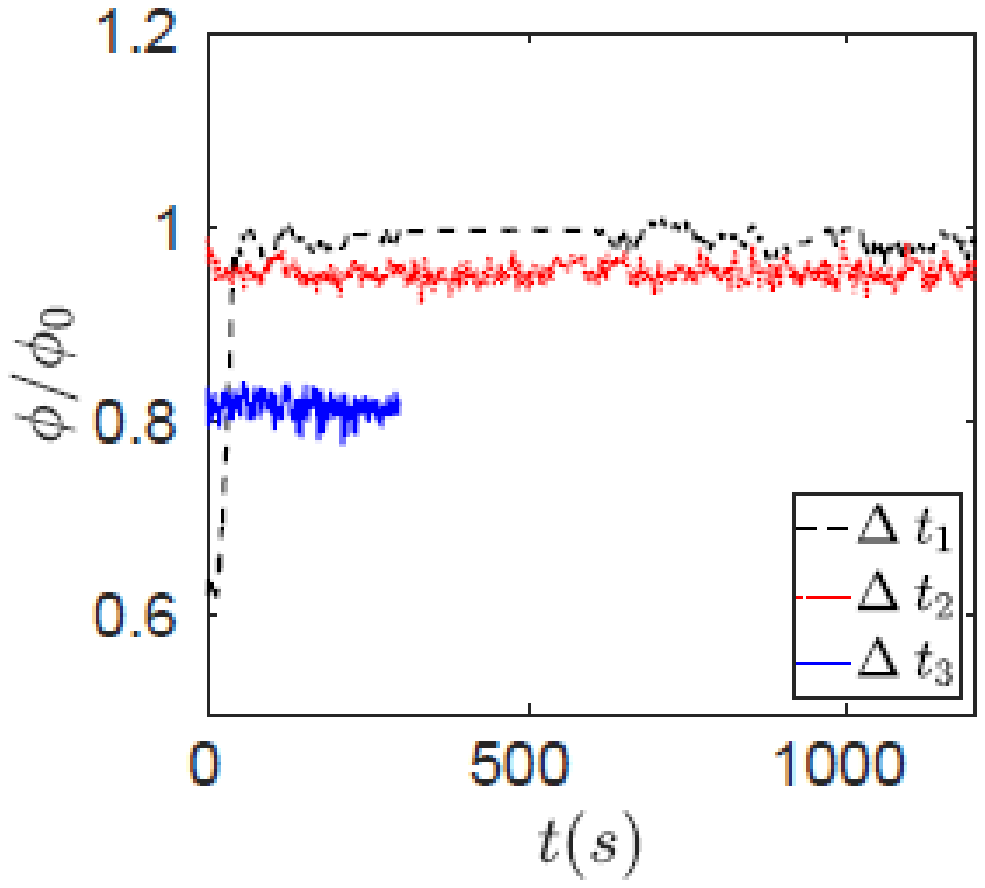}\\
		(b)
    \end{center}
   \end{minipage}
	\\
   \begin{minipage}[c]{0.35\linewidth}
    \begin{center}
      \includegraphics[width=.99\linewidth]{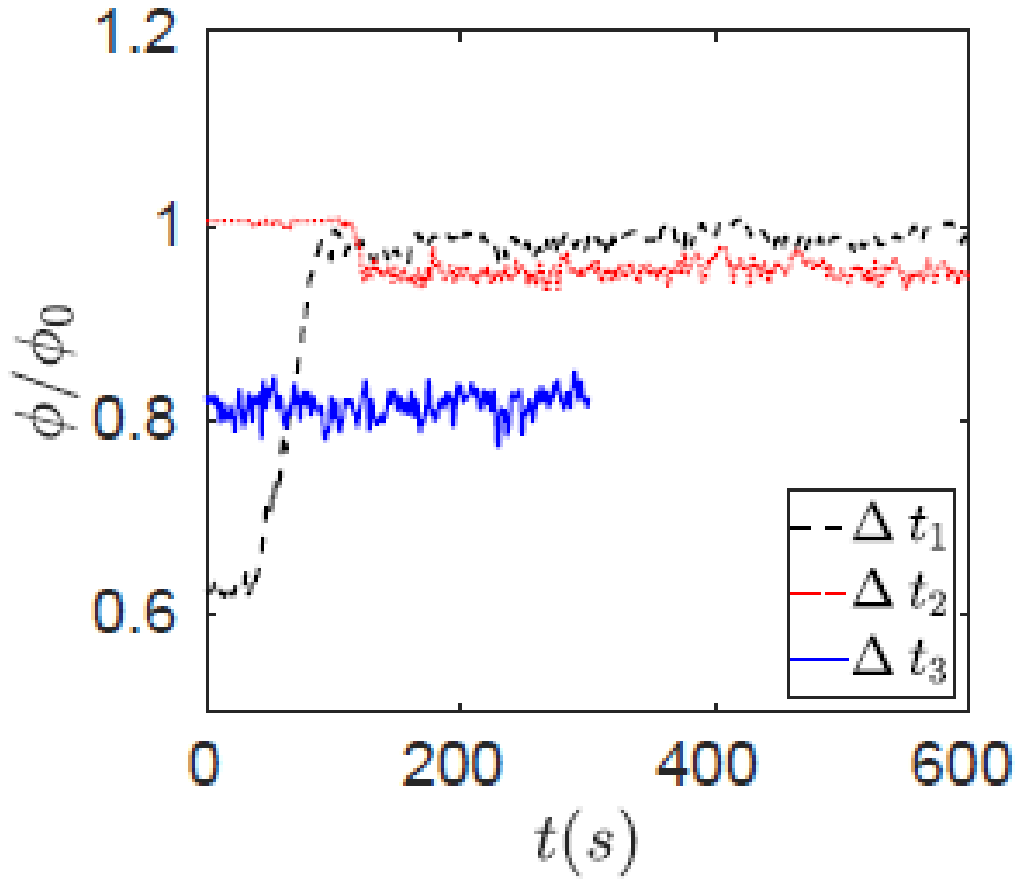}\\
		(c)
    \end{center}
   \end{minipage}
   \begin{minipage}[c]{0.35\linewidth}
    \begin{center}
      \includegraphics[width=.99\linewidth]{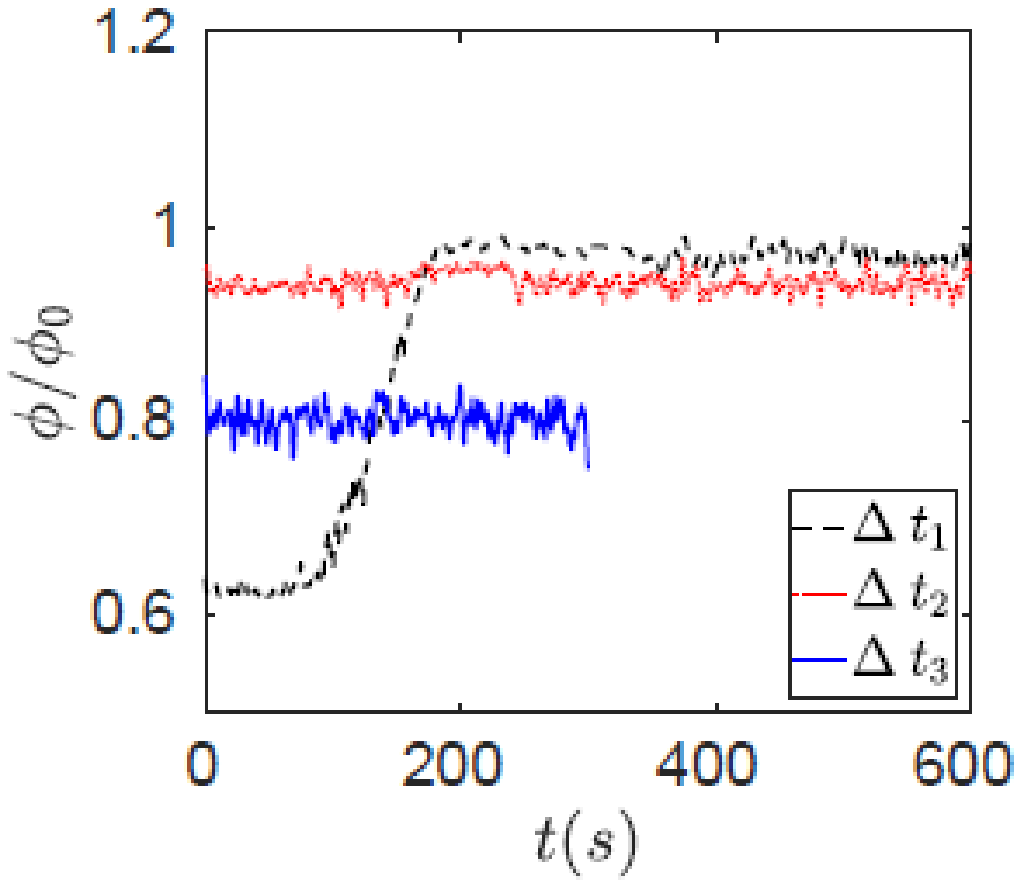}\\
		(d)
    \end{center}
   \end{minipage}

\caption{Evolution of the mean particle fraction along the different intervals. Figures (a), (b), (c) and (d) correspond to tests 1, 8, 11 and 16, respectively. The corresponding intervals are listed in the key.}
	\label{fig:phi}
\end{figure}

Based on image processing of all experiments, we identified the different structures and their respective durations, which are shown in Tab. \ref{table:table2}. In addition, we measured the bed heights and obtained the mean particle fraction, computed as the bed volume divided by that of individual beads. Figure \ref{fig:phi} shows the mean particle fraction normalized by the value corresponding to incipient fluidization, $\phi / \phi_0$, as a function of time for intervals $\Delta t_1$, $\Delta t_2$ and $\Delta t_3$. Figures \ref{fig:phi}(a), \ref{fig:phi}(b), \ref{fig:phi}(c) and \ref{fig:phi}(d) correspond to tests 1, 8, 11 and 16 (Tab. \ref{table:table2}), respectively, and represent of the ensemble of behaviors observed in our tests with species 1 and 3 in terms of occurrence of crystallization and jamming.

We observe first that in our tests the particle fractions of crystallized and jammed beds were equal or lower than that of the initial bed, meaning that beds are expanded with respect to incipient fluidization conditions. Then, we observe that, for beds that crystallized, the jammed state maintains approximately the same particle fraction, meaning that the grains basically stopped their microscopic motion when compared with the crystallized condition. Finally, when the water velocity is increased and the bed re-fluidized, the particle fraction decreases. This is the general picture for most tests, but some variations happened. For example, in some few tests, such as test 8, jamming did not occur, and, for that reason, we can observe in Fig. \ref{fig:phi}(b) that $\phi$ oscillates and attains values lower than that of crystallization. In other few cases, such as in test 11, crystallization did not occur but jamming appeared during part of $\Delta t_2$. For these cases, Fig. \ref{fig:phi}(c) shows that the bed attains a higher value of  $\phi$ during jamming. For other few cases, neither crystallization nor jamming occurred (Fig. \ref{fig:phi}(d)), and the value of $\phi$ decreased with the increase of the water velocity. Finally, the most strong variation during $\Delta t_3$ was the appearance of a volcano-like structure for species 2. However, given the radial variation of this kind of structure, we understand that the mean void fraction is not a pertinent parameter and we did not compute it. Instead, we analyze that structure in terms of intensity (next).

\begin{figure}[h!]
   \begin{minipage}[c]{0.35\linewidth}
    \begin{center}
     \includegraphics[width=.99\linewidth]{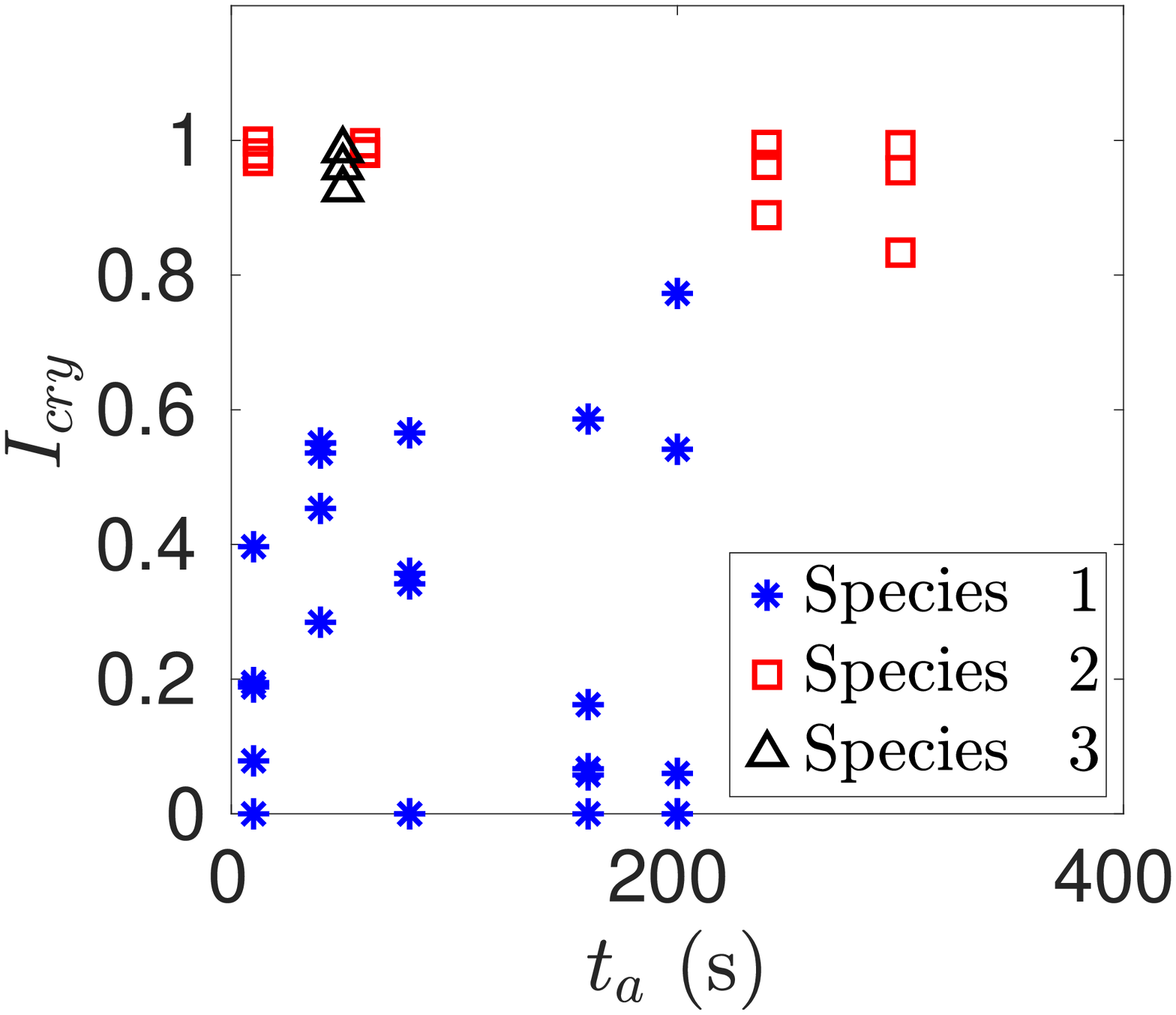}\\
		(a)
    \end{center}
   \end{minipage}
   \begin{minipage}[c]{0.35\linewidth}
    \begin{center}
      \includegraphics[width=.99\linewidth]{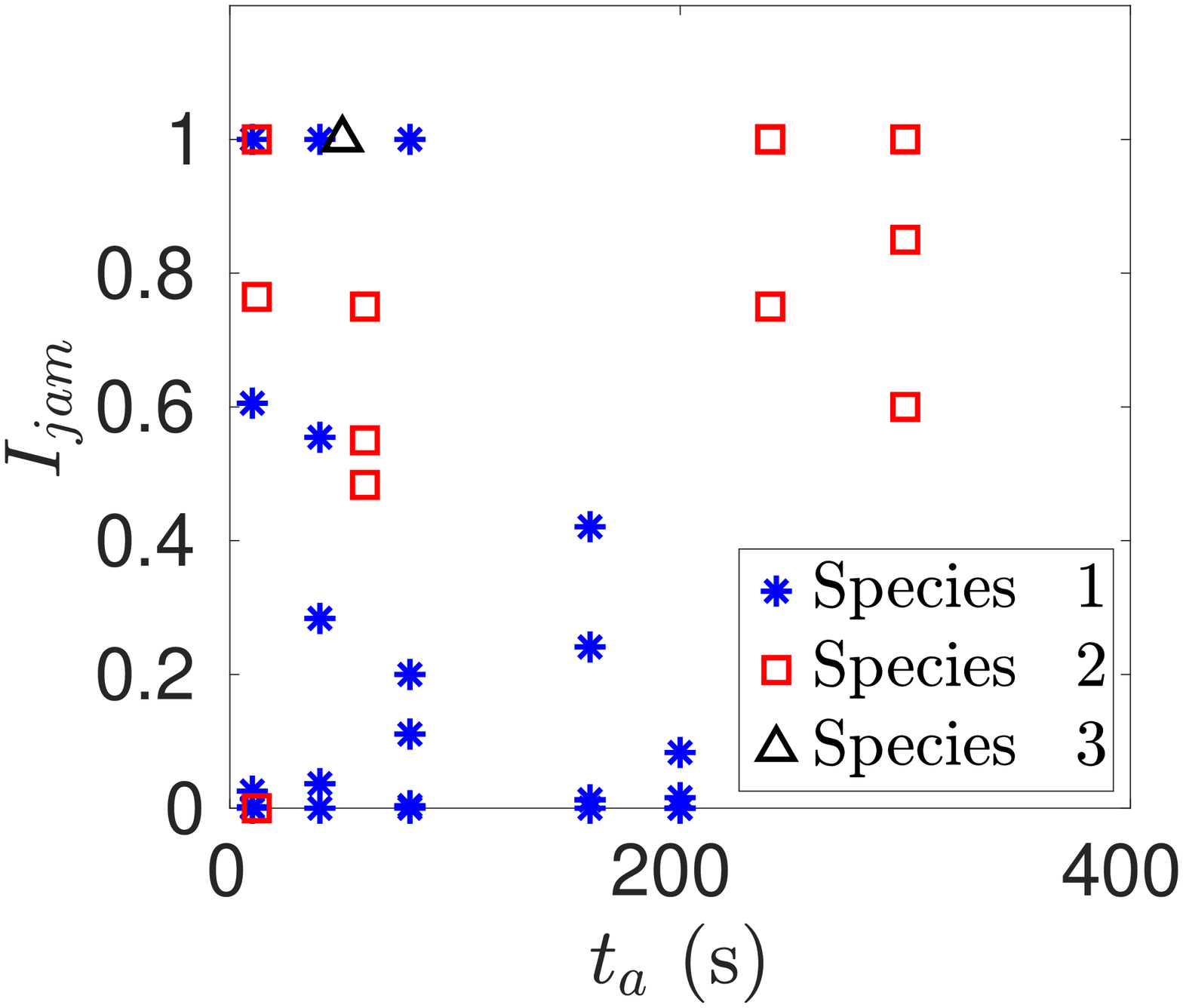}\\
		(b)
    \end{center}
   \end{minipage}
	\caption{Intensities of (a) crystallization and (b) jamming as functions of the deceleration time $t_a$.}
	\label{fig_intensities}
\end{figure}

As in some tests crystallization and Jamming appeared only once and lasted until the end of intervals $\Delta t_1$ and $\Delta t_2$, respectively, while in some others they appeared and disappeared more than once, and in still other tests they did not appear, we measured the total duration of crystallization and jamming and computed their intensities $I_{cry}$ and $I_{jam}$, respectively. Because in their experiments Goldman and Swinney \cite{Goldman} found that those states depend on the deceleration rate, we investigate next their variation with $t_a$, Figs. \ref{fig_intensities}(a) and \ref{fig_intensities}(b) showing, respectively, $I_{cry}$ and $I_{jam}$ as functions of the deceleration time $t_a$. It is remarkable how both the crystallization and jamming intensities do not depend on the deceleration rate, but rather on the grain type: beds consisting of larger and lighter grains tend to be crystallized and jammed. This is in clear contrast with the findings of Goldman and Swinney \cite{Goldman}. However, while they used only one grain type and did not investigate the effect of grain size and weight, our beds were much narrower than theirs ($D/d$ $\leq$ 4.2 in our case and $D/d$ $\sim$ 100 in Goldman and Swinney \cite{Goldman}). Therefore, in the case of very narrow tubes (at least $D/d$ $\leq$ 5), the grain characteristics determines the behavior of the bed under de-fluidizing and fluidizing conditions, different from less narrow beds. This is an important characteristic of highly confined beds that had been unknown until now.

\begin{figure}[h!]
   \begin{minipage}[c]{0.35\linewidth}
    \begin{center}
     \includegraphics[width=.99\linewidth]{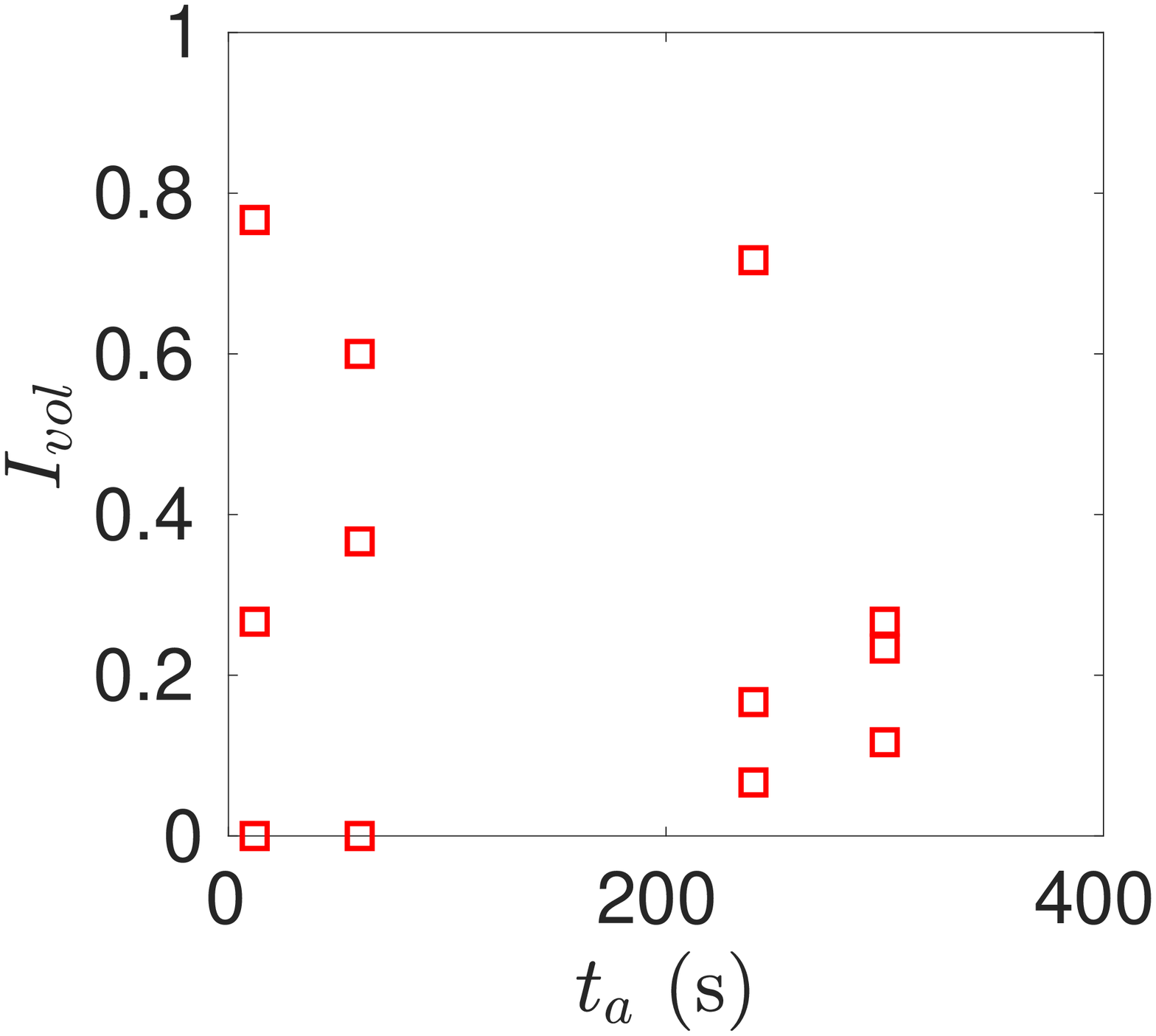}\\
		(a)
    \end{center}
   \end{minipage}
   \begin{minipage}[c]{0.35\linewidth}
    \begin{center}
      \includegraphics[width=.99\linewidth]{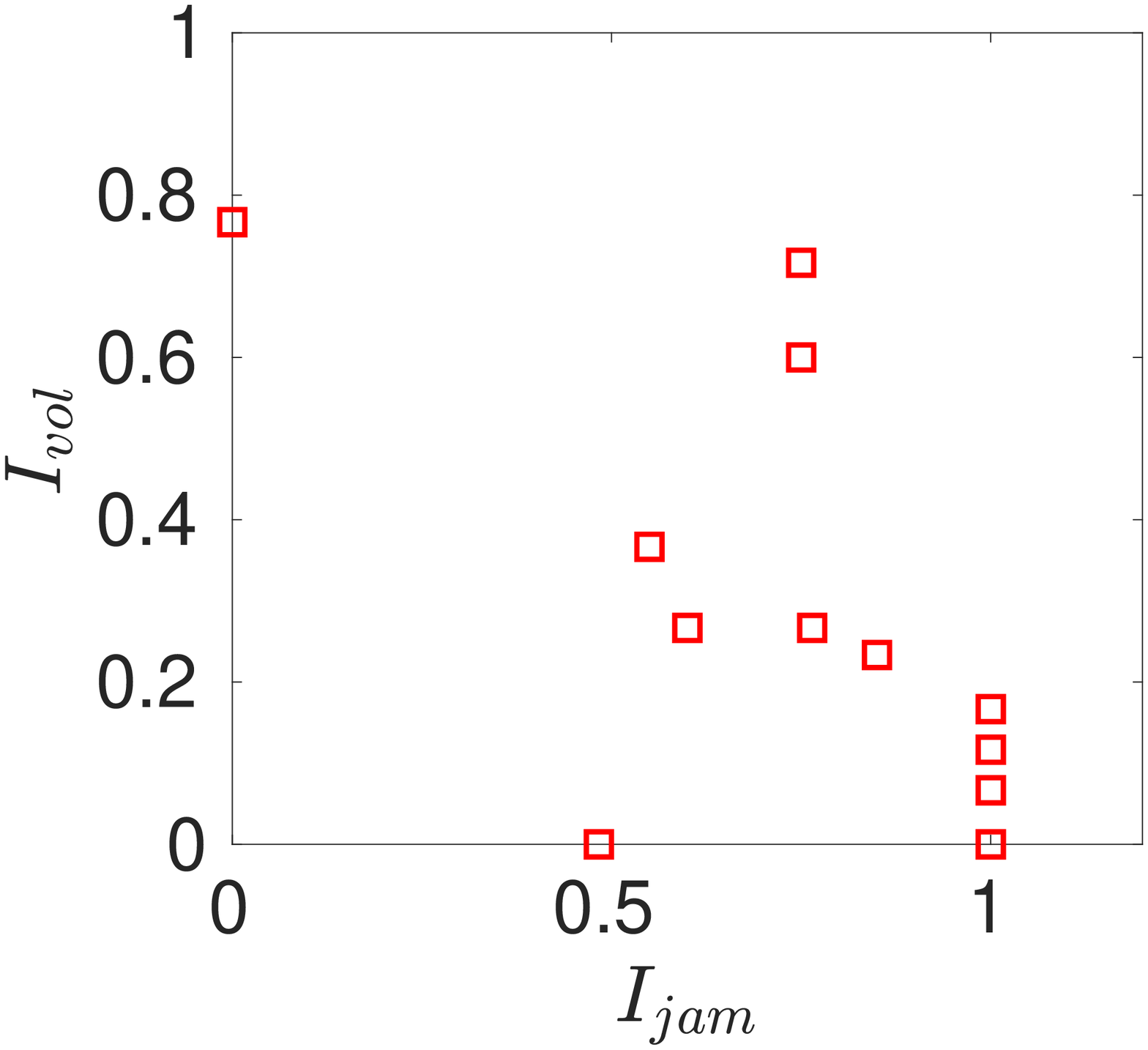}\\
		(b)
    \end{center}
   \end{minipage}
	\caption{Intensity of the volcano-like structure as function of (a) the deceleration time $t_a$ and (b) intensity of jamming $I_{jam}$.}
	\label{fig_intensities_vol}
\end{figure}

Finally, we computed the intensity of the volcano-like structure appearing for species 2, $I_{vol}$, and analyzed its behavior a function of the deceleration time $t_a$, as shown in Fig. \ref{fig_intensities_vol}(a), but also of the intensity of jamming $I_{jam}$ (once jamming occured just before the volcano crystallization), as shown in Fig. \ref{fig_intensities_vol}(b). We could find no clear dependence on neither $t_a$ nor $I_{jam}$. The transition from jamming to the volcano structure is also a characteristic of highly confined beds that had not been reported until now, but it rests to be further investigated.

\subsection{Microscopic observations}
\label{subsec:microscopic}

In addition to identifying bed structures, we tracked individual grains along movie frames with numerical scripts based on Kelley and Ouellette \cite{Kelley} and Houssais et al. \cite{Houssais_1}. Because the bed was three dimensional, only grains in contact with the tube wall within the field of view of the camera could be tracked. Therefore, measurements were based on a frontal view of a cylindrical plane, which we associated with a Cartesian coordinate system. With the position of tracked grains, we computed their trajectories and obtained two-dimensional displacements and velocities. In order to further investigate the crystallization and jamming processes, we computed the $x$ and $y$ (Fig. \ref{fig:1}) components of the instantaneous velocity of each grain, $U_p$ and $V_p$, respectively, of velocity fluctuations, $u_p$ and $v_p$, respectively, and the two-dimensional granular temperature $\theta$ as in Eq. \ref{Eq:grain_temp}. The velocity fluctuations were computed as the deviation of the instantaneous velocity of each grain from the average value for the ensemble of grains.

\begin{equation}
\theta \,=\, \frac{1}{2} \left( u_p^2+v_p^2 \right)
\label{Eq:grain_temp}
\end{equation}

\begin{figure}[h!]
   \begin{minipage}[c]{0.35\linewidth}
    \begin{center}
     \includegraphics[width=.99\linewidth]{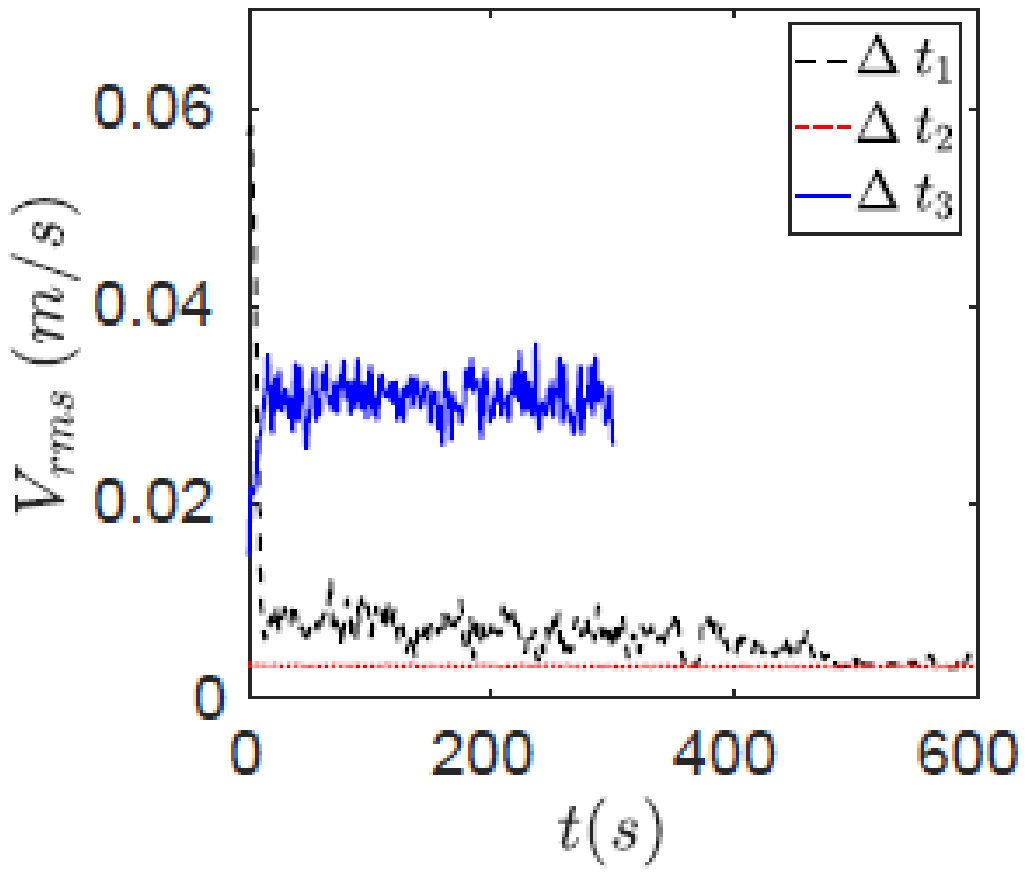}\\
		(a)
    \end{center}
   \end{minipage}
   \begin{minipage}[c]{0.35\linewidth}
    \begin{center}
      \includegraphics[width=.99\linewidth]{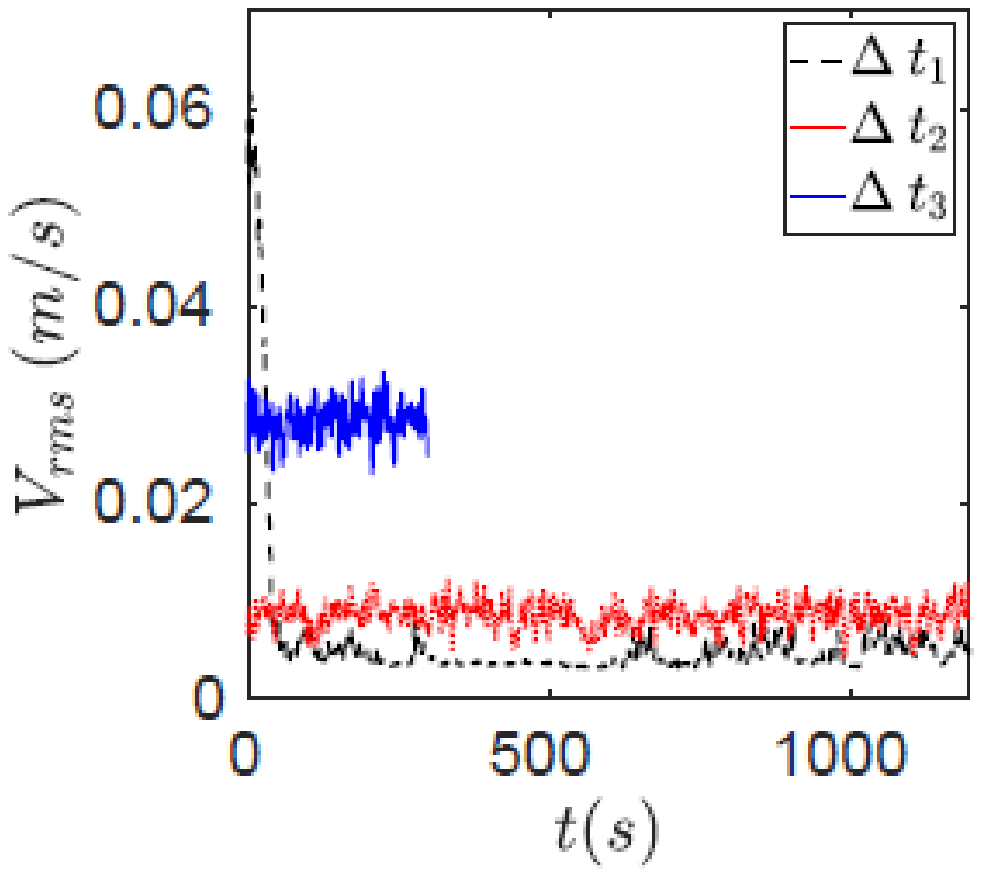}\\
		(b)
    \end{center}
   \end{minipage}
	\\
   \begin{minipage}[c]{0.35\linewidth}
    \begin{center}
      \includegraphics[width=.99\linewidth]{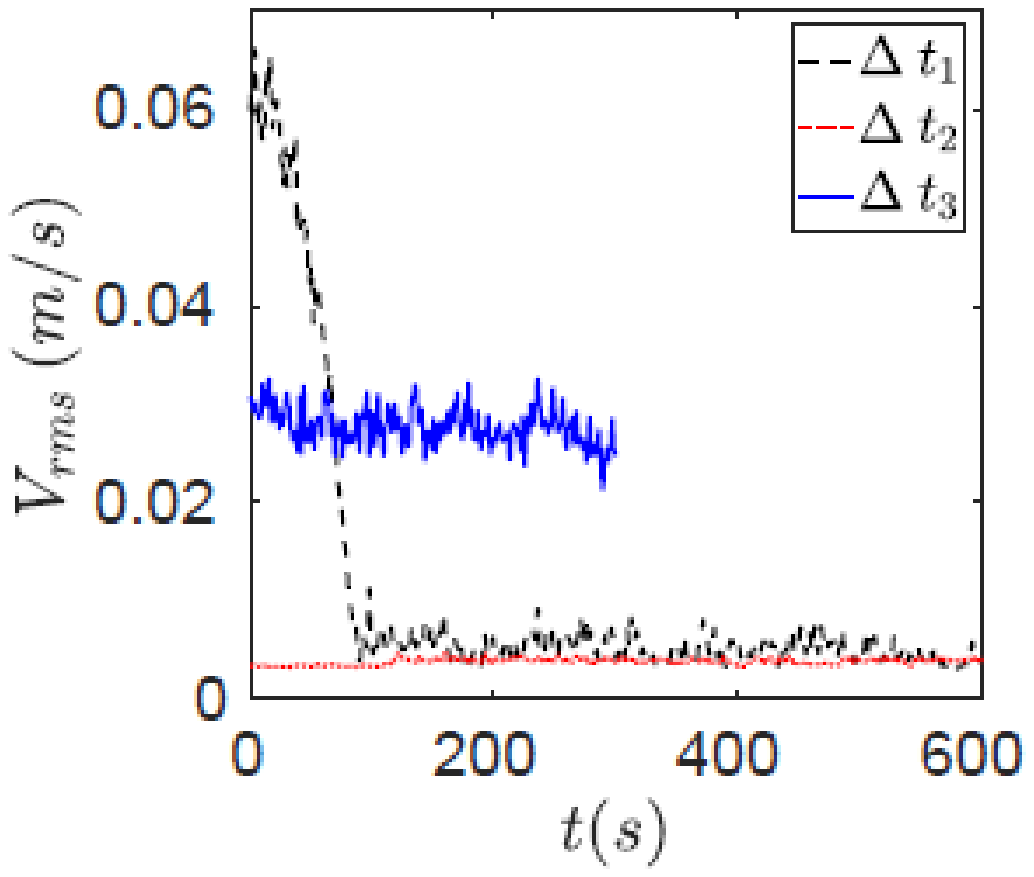}\\
		(c)
    \end{center}
   \end{minipage}
   \begin{minipage}[c]{0.35\linewidth}
    \begin{center}
      \includegraphics[width=.99\linewidth]{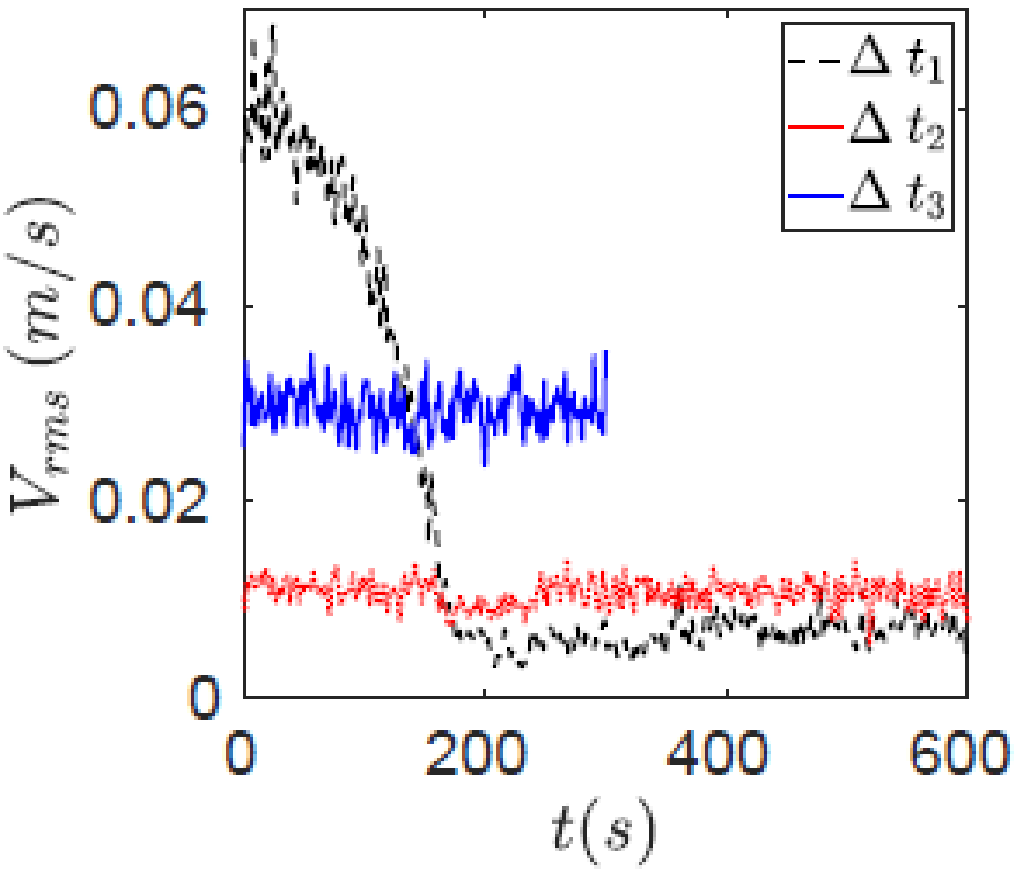}\\
		(d)
    \end{center}
   \end{minipage}
\caption{Evolution of the rms average of the norm of the velocity for all particles along the different intervals. Figures (a), (b), (c) and (d) correspond to tests 1, 8, 11 and 16, respectively. The corresponding intervals are listed in the key.}
	\label{fig:rms}
\end{figure}

\noindent In addition, we computed a rms average of the norm of the velocity (in the $xy$ plane), as in Eq. \ref{Eq:rms_vel},

\begin{equation}
V_{rms} \,=\, \sqrt{\sum_{i=1}^N \frac{1}{N} \left( U_p \right) ^2 + \sum_{i=1}^N \frac{1}{N} \left( V_p \right) ^2}
\label{Eq:rms_vel}
\end{equation}

\noindent where $V_{rms}$ is the instantaneous rms average for the ensemble of grains, $i$ refers to the $i^{th}$ grain, and $N$ is the number of considered grains (appearing in the image). Figures \ref{fig:rms}(a), \ref{fig:rms}(b), \ref{fig:rms}(c) and \ref{fig:rms}(d) present $V_{rms}$ for tests 1, 8, 11 and 16, respectively.

The rms average of the norm of the velocity indicates the degree of microscopic motion of grains, as can be observed in Fig. \ref{fig:rms}. At the beginning, when bed is fluidized, we observe large values of $V_{rms}$, which in their turn fluctuate around a mean value. For the initial velocity $U_a$, the mean value of $V_{rms}$ is approximately 0.06 m/s. At $U_b$, values are much smaller, around 0.006 m/s, but still with fluctuations around a mean value, and, when the bed crystallizes, fluctuations of $V_{rms}$ are considerable reduced. When the water velocity is increased to $U_c$, $V_{rms}$ is still reduced to values around 0.003 m/s if jamming occurs, and fluctuations of $V_{rms}$ virtually disappear. In cases where fluidization occurs, both the value of $V_{rms}$ and its fluctuations increase. When the water velocity reaches $U_d$, the bed fluidizes and $V_{rms}$ fluctuates around 0.03 m/s.

\begin{figure}[h!]
   \begin{minipage}[c]{0.49\linewidth}
    \begin{center}
     \includegraphics[width=.99\linewidth]{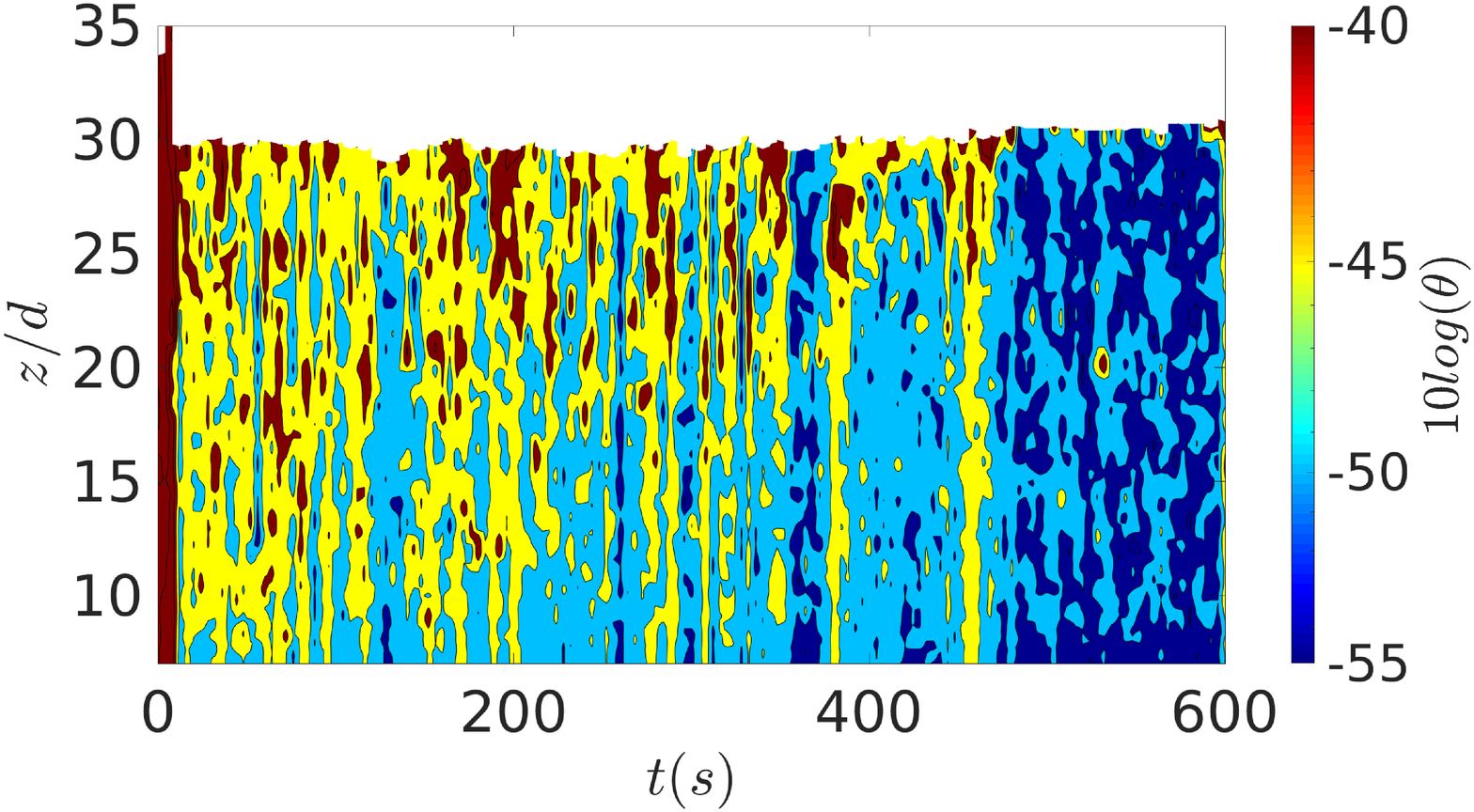}\\
		(a)
    \end{center}
   \end{minipage}
   \begin{minipage}[c]{0.49\linewidth}
    \begin{center}
      \includegraphics[width=.99\linewidth]{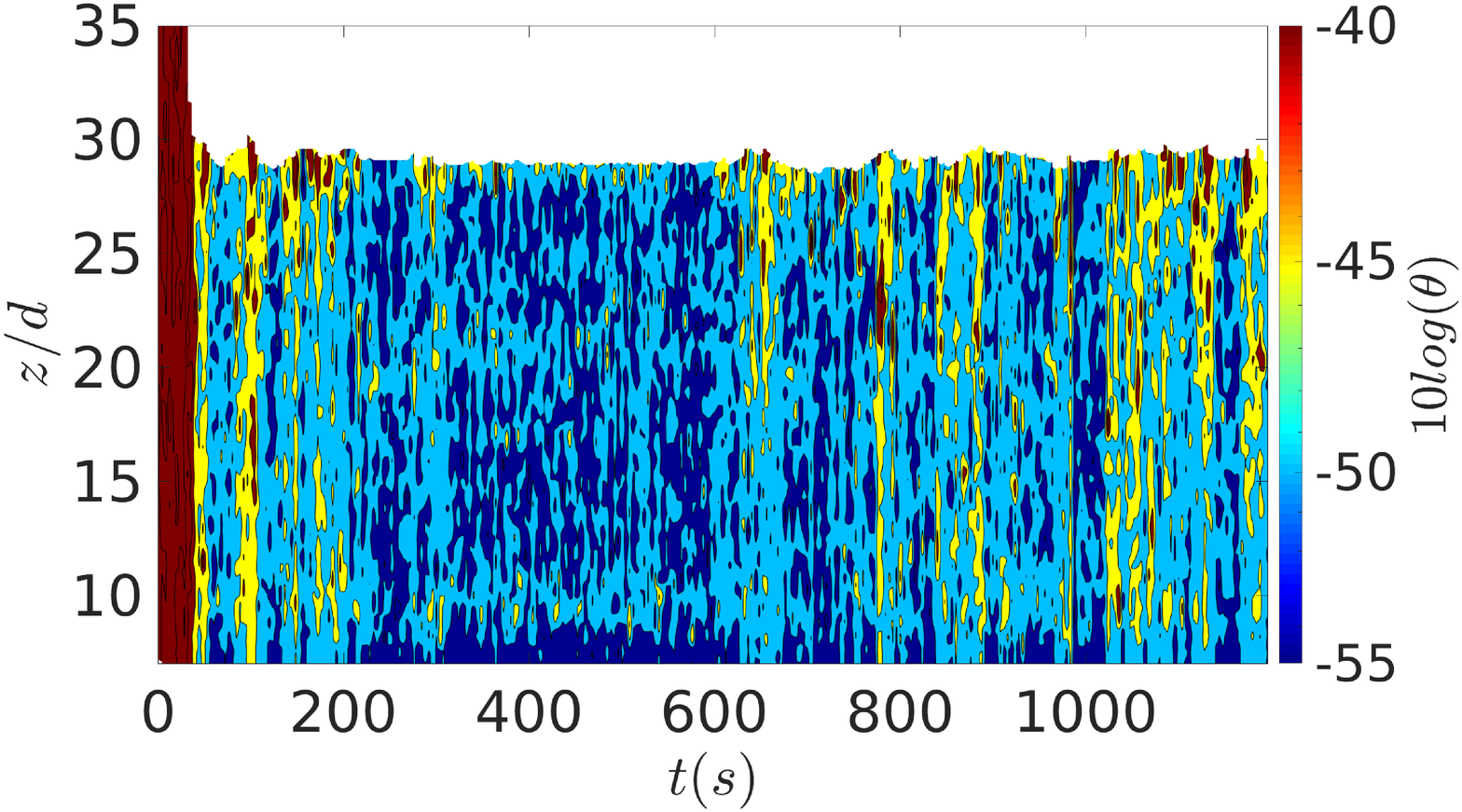}\\
		(b)
    \end{center}
   \end{minipage}
	\\
   \begin{minipage}[c]{0.49\linewidth}
    \begin{center}
      \includegraphics[width=.99\linewidth]{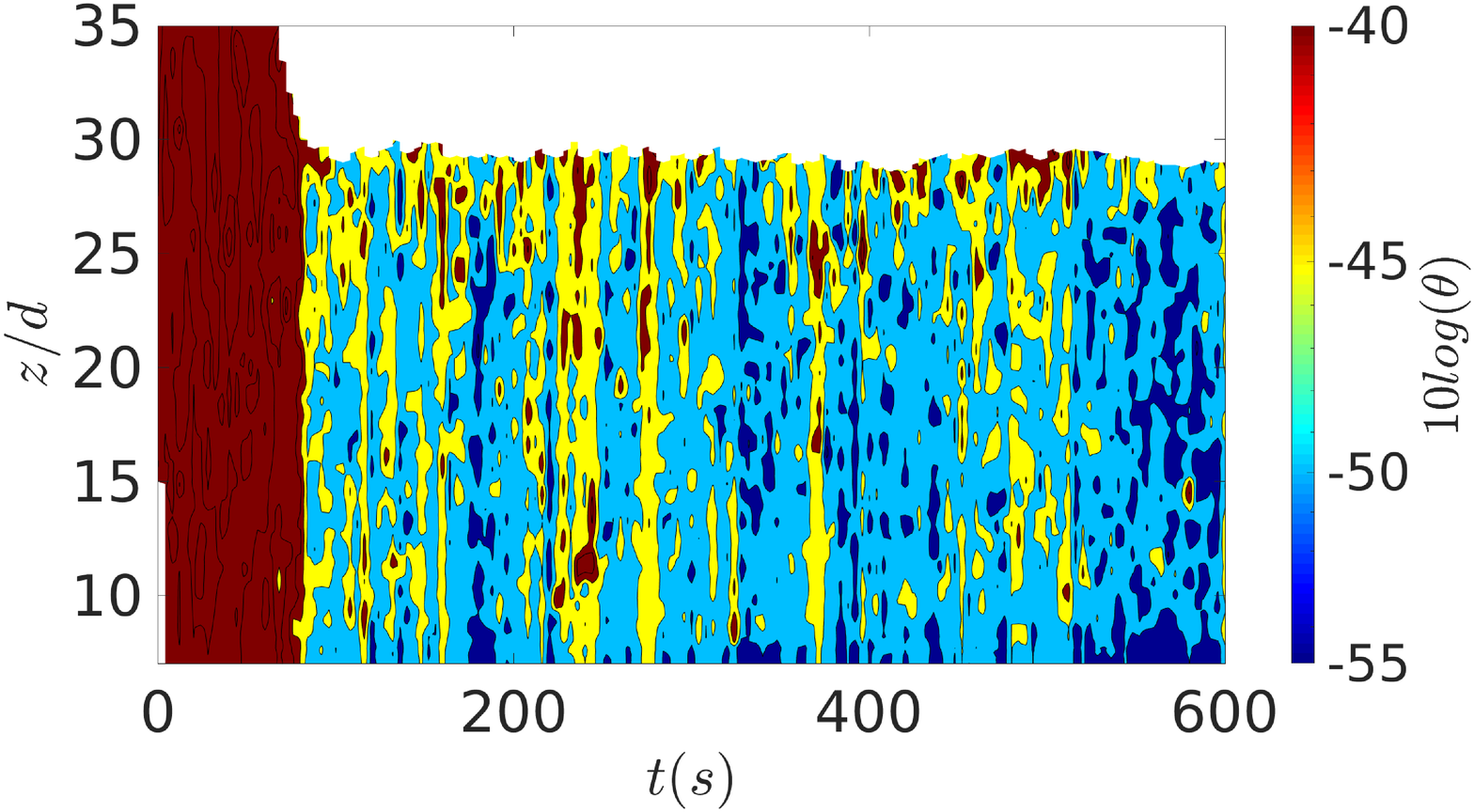}\\
		(c)
    \end{center}
   \end{minipage}
   \begin{minipage}[c]{0.49\linewidth}
    \begin{center}
      \includegraphics[width=.99\linewidth]{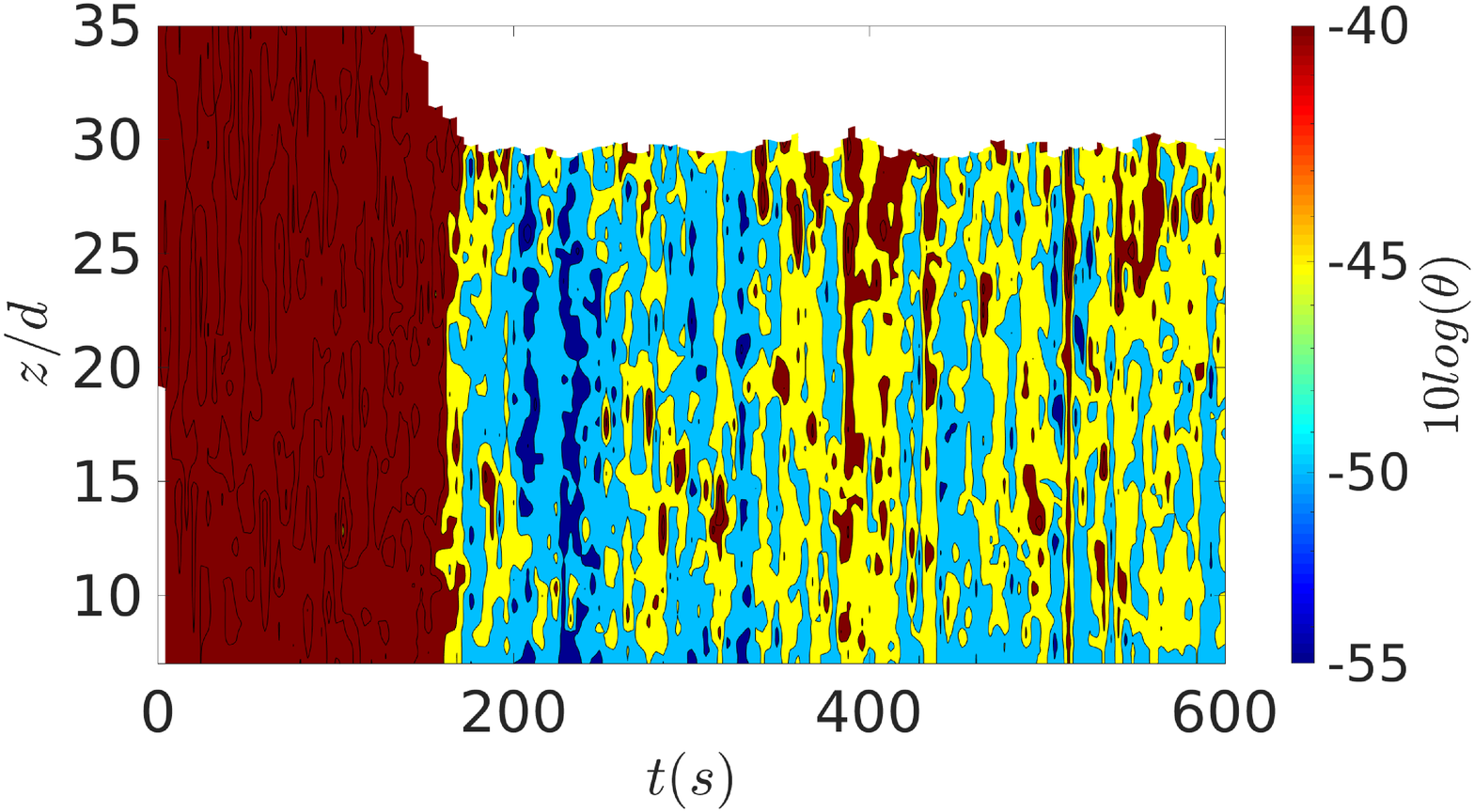}\\
		(d)
    \end{center}
   \end{minipage}
\caption{Spatio-temporal diagrams of cross-sectional averages of the granular temperature during $\Delta t_1$. Figures (a), (b), (c) and (d) correspond to tests 1, 8, 11 and 16, respectively. The corresponding temperatures are listed in the key.}
	\label{fig:gran_temp_cry}
\end{figure}

Based on the two-dimensional granular temperatures (Eq. \ref{Eq:grain_temp}), we estimated the instantaneous values of cross-sectional averages by computing horizontal averages of $\theta$ for each frame. Figures \ref{fig:gran_temp_cry}(a), \ref{fig:gran_temp_cry}(b), \ref{fig:gran_temp_cry}(c) and \ref{fig:gran_temp_cry}(d) show spatio-temporal diagrams of cross-sectional averages of the granular temperature for tests 1, 8, 11 and 16, respectively, during $\Delta t_1$, where crystallization occurred for some tests. We plotted 10$\log \theta$ instead of $\theta$ in order to accentuate differences. For all cases, we note initially a rapid decrease in $\theta$, during the deceleration, and then lower values for the cases where crystallization occurred, Figs. \ref{fig:gran_temp_cry}(a) and \ref{fig:gran_temp_cry}(b). Of particular interest is the fact that $\theta$ reaches lower values first at the bottom of the bed, where crystallization appears first, and those lower values spread upwards as crystallization propagates toward the top, as can be seen in Figs. \ref{fig:gran_temp_cry}(a) and \ref{fig:gran_temp_cry}(b) considering their respective $t_{cry}$. Animations showing the time evolution of the distribution of $\theta$ within the bed are available as Supplementary Material and in Mendeley Data \cite{Supplemental3}.

\begin{figure}[h!]
   \begin{minipage}[c]{0.49\linewidth}
    \begin{center}
     \includegraphics[width=.99\linewidth]{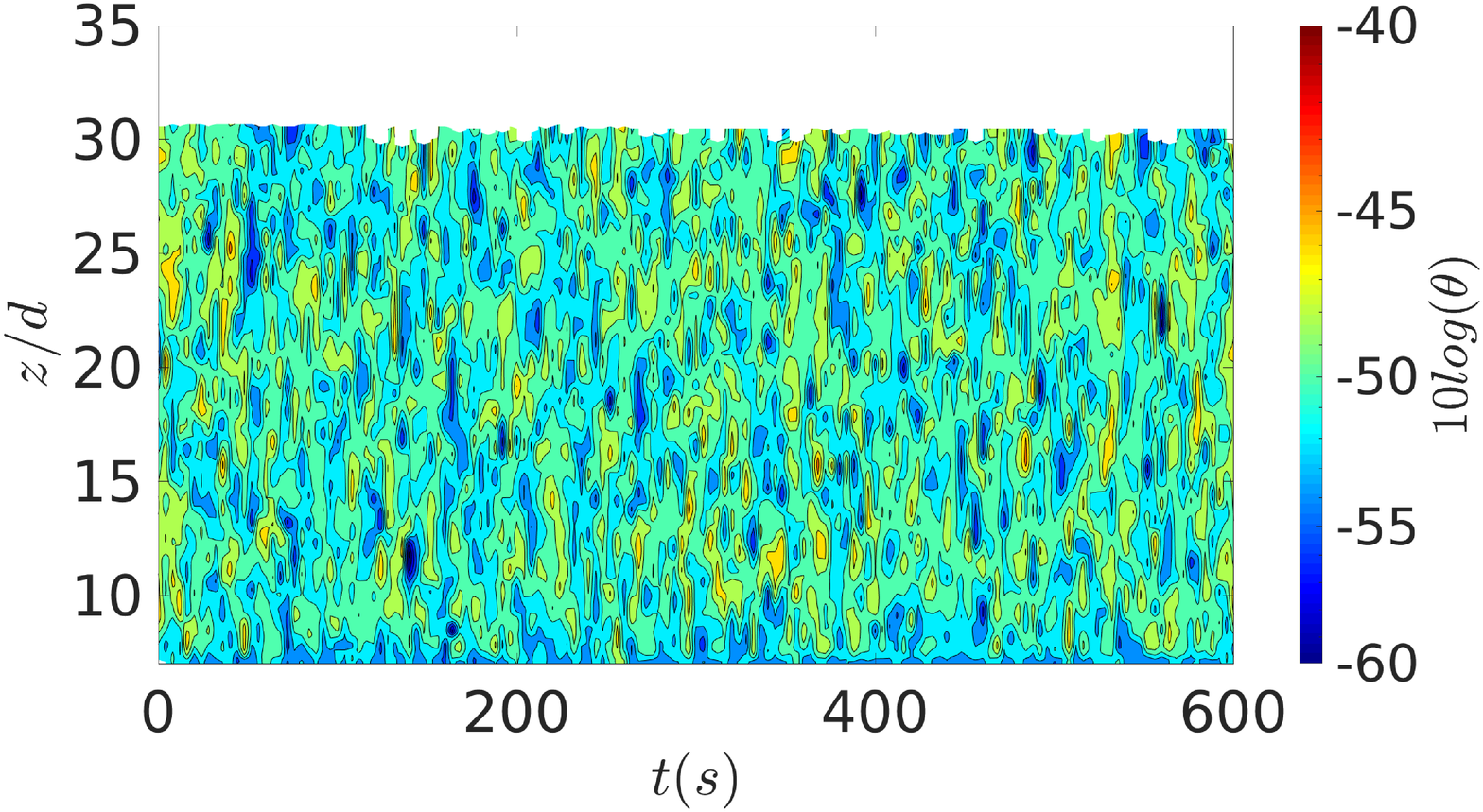}\\
		(a)
    \end{center}
   \end{minipage}
   \begin{minipage}[c]{0.49\linewidth}
    \begin{center}
      \includegraphics[width=.99\linewidth]{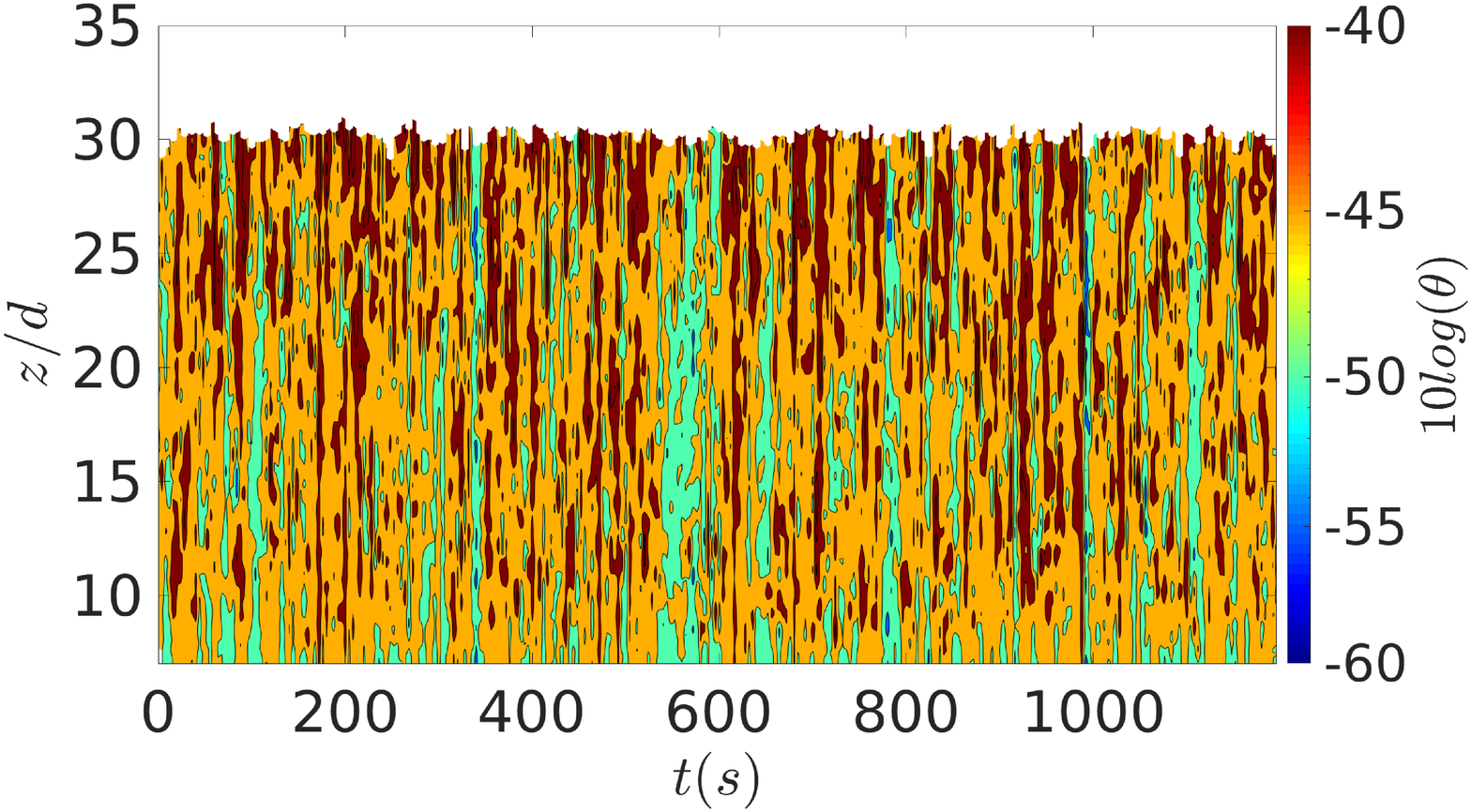}\\
		(b)
    \end{center}
   \end{minipage}
	\\
   \begin{minipage}[c]{0.49\linewidth}
    \begin{center}
      \includegraphics[width=.99\linewidth]{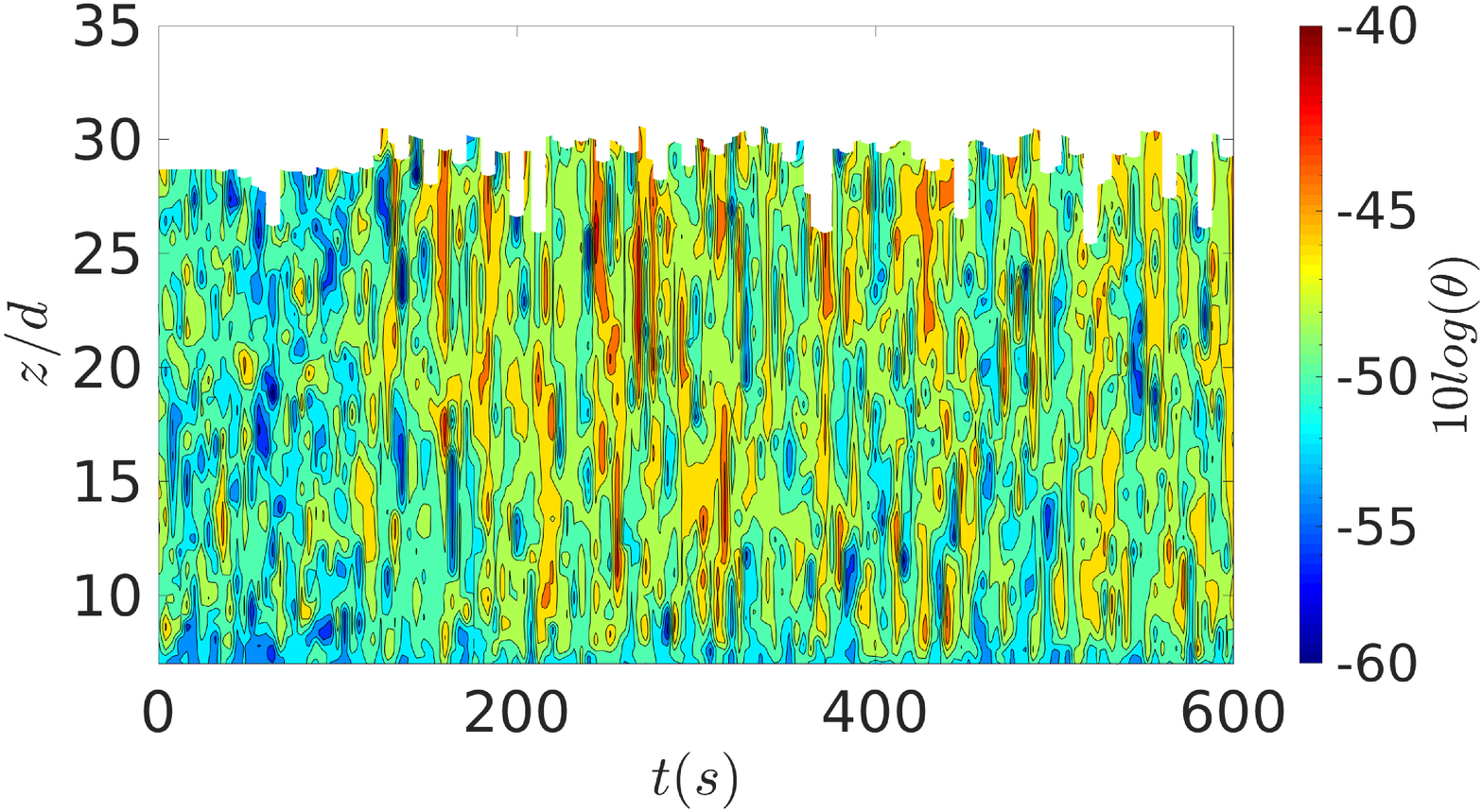}\\
		(c)
    \end{center}
   \end{minipage}
   \begin{minipage}[c]{0.49\linewidth}
    \begin{center}
      \includegraphics[width=.99\linewidth]{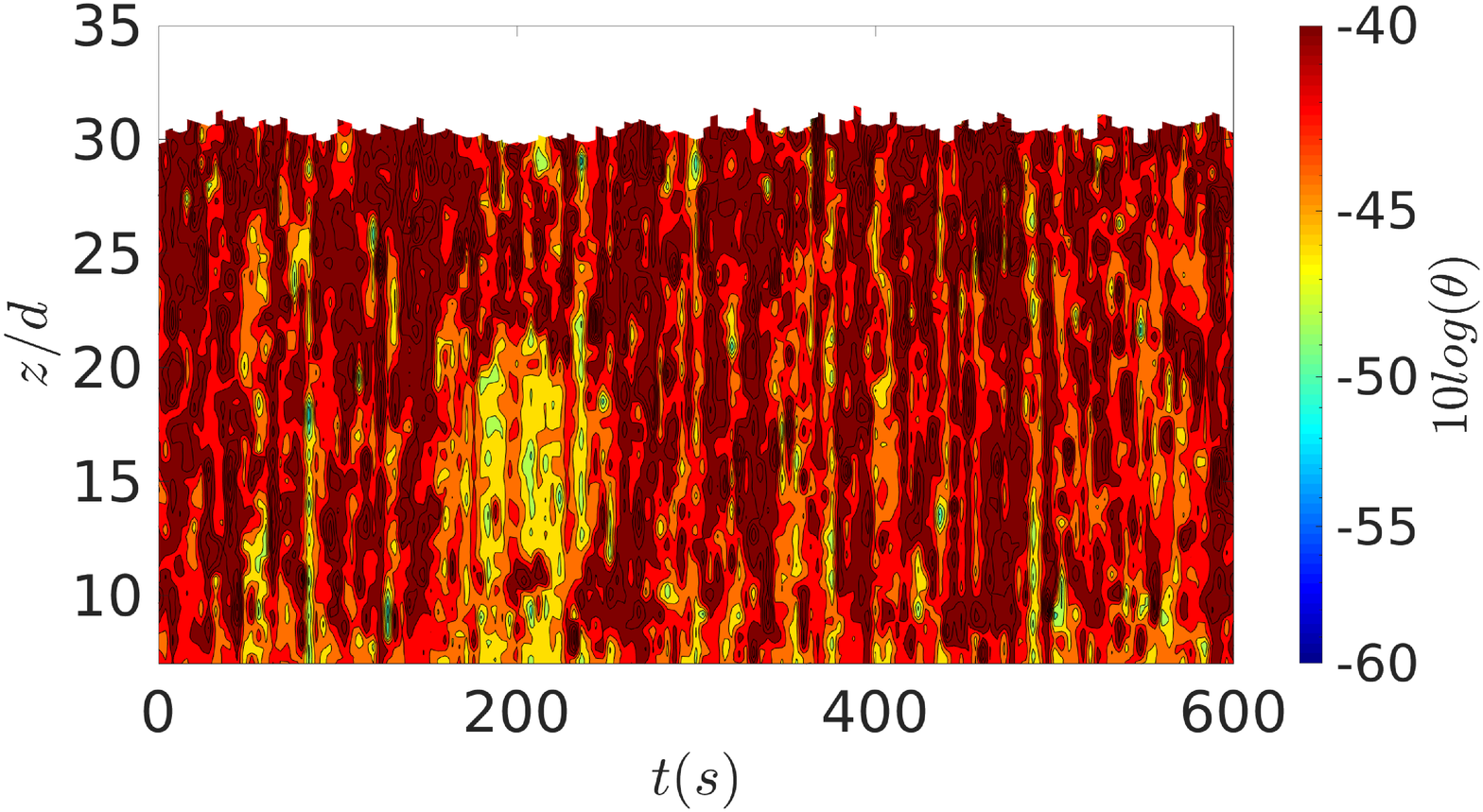}\\
		(d)
    \end{center}
   \end{minipage}
\caption{Spatio-temporal diagrams of cross-sectional averages of the granular temperature during $\Delta t_2$. Figures (a), (b), (c) and (d) correspond to tests 1, 8, 11 and 16, respectively. The corresponding temperatures are listed in the key.}
	\label{fig:gran_temp_jam}
\end{figure}

Figures \ref{fig:gran_temp_jam}(a), \ref{fig:gran_temp_jam}(b), \ref{fig:gran_temp_jam}(c) and \ref{fig:gran_temp_jam}(d) show spatio-temporal diagrams of cross-sectional averages of the granular temperature for tests 1, 8, 11 and 16, respectively, during $\Delta t_2$, where jamming occurred for some tests. We note the much lower values of $\theta$ during jammings, reflecting small level of microscopic motion in the jammed state.

We found that for velocities slightly higher than $U_{if}$ the bed can crystallize, being organized in a static lattice of high compactness without macroscopic motion. While we found that crystal structures are initially localized and grow along time, in a similar manner as happens in phase transitions, we did not find any dependence on the decreasing rate, as found by Goldman and Swinney \cite{Goldman}. According to Goldman and Swinney \cite{Goldman}, both the local growth and rate dependence would have similarities with glass formation in fluids, where the role of fluid velocity is associated to that of temperature. Jamming usually appears when the crystallized bed is forced upwards by submitting it to a fluid velocity slightly higher than that for crystallization, with grains that are already packed together being forced against each other. Given the high confinement of narrow beds, they become in closer contact with each other and with the tube wall and the microscopic motion vanishes. Previous works have associated jamming in grains with glass formation \cite{Liu2, OHern}, where a glass state is formed by cooling a liquid below its freezing point (supercooled liquid) until solidification (glass) occurs, usually obtained by using high cooling rates, which avoid crystallization below the freezing point and make the liquid viscosity increase drastically so that it becomes a glass below the glass transition temperature. Therefore, the similarity between glass and jamming would be particularly interesting in cases where crystallization does not occur but jamming does; however, this remains to be investigated further.

Finally, crystallization and jamming raise questions concerning the fluidization of narrow beds. In most of cases, for velocities above that for incipient fluidization, our experiments showed that, after being fluidized, beds crystallized and jammed, the latter having static conditions even at the grain scale. Therefore, in the case highly confined beds, considerations about crystallization and jamming should be taken into account in determining fluidization conditions.

\section{CONCLUSIONS}
\label{sec:conclusions}

In this paper we investigated the crystallization and jamming happening at liquid velocities higher than that assuring incipient fluidization in solid-liquid fluidized beds (SLFBs). We were interested in the case of very narrow tubes (ratio between the tube and grain diameters smaller than 5), for which confinement effects are high, and we varied the grain types, water velocities, resting times, and velocity decelerations. We found that, after a decelerating water flow that reached a velocity still higher than that necessary for bed fluidization, grains crystallized in most cases, i.e., became organized in lattice structures occupying the entire tube cross section, where they were trapped though with small fluctuations (microscopic motion). After a certain time, when the liquid velocity was slightly increased, jamming occurred, with microscopic motion disappearing.

We found that both the crystallization and jamming intensities do not depend on the deceleration rate, but rather on the grain type. These findings are in clear contrast with those of Goldman and Swinney \cite{Goldman}, who found that the final state depends on the decreasing rate. Our results show that, for very narrow beds, the grain characteristics determines the bed behavior under defluidizing and slight fluidizing conditions. This is an important characteristic of highly confined beds that had been unknown until now.

Finally, we showed that the crystal structures are initially localized and grow along time, in a similar manner as happens in phase transitions, and that a second structure may appear in some cases with a volcano-like form. However, the conditions for the appearance of the latter were not identified and need to be investigated further. Our findings raise the question of fluidization conditions, which should take into consideration crystallization and jamming in the case of highly confined beds.

\section*{SUPPLEMENTARY MATERIAL}
See Supplementary Material for microscopy images of the used grains, instantaneous snapshots of particle positions for tests 1, 8, 11 and 16, movies showing the evolution of a fluidized bed, and animations showing the distribution of granular temperature.

\section*{DATA AVAILABILITY}
The data that support the findings of this study are openly available in Mendeley Data \cite{Supplemental3} at http://dx.doi.org/10.17632/j28vhs37n8.1.

\begin{acknowledgments}
Fernando David C\'u\~nez is grateful to FAPESP (Grant Nos. 2016/18189-0 and 2018/23838-3), and Erick Franklin would like to express his gratitude to FAPESP (Grant No. 2018/14981-7) and to CNPq (Grant No. 400284/2016-2) for the financial support they provided.
\end{acknowledgments}

\bibliography{references1}

\end{document}